\documentclass[prb,showpacs]{revtex4}
\usepackage{amssymb}
\usepackage{graphicx}
\usepackage{amsmath}

\input{tcilatex}

\begin{document}

\title{Spectral Density Functionals for Electronic Structure Calculations}
\author{S. Y. Savrasov}
\affiliation{Department of Physics, New Jersey Institute of Technology, Newark, NJ 07102}
\author{G. Kotliar}
\affiliation{Department of Physics and Astronomy and Center for Condensed Matter Theory,
Rutgers University, Piscataway, NJ 08854}

\begin{abstract}
We introduce a spectral density functional theory which can be used to
compute energetics and spectra of real strongly--correlated materials using
methods, algorithms and computer programs of the electronic structure theory
of solids. The approach considers the total free energy of a system as a
functional of a local electronic Green function which is probed in the
region of interest. Since we have a variety of notions of locality in our
formulation, our method is manifestly basis--set dependent. However, it
produces the exact total energy and local excitational spectrum provided
that the exact functional is extremized. The self--energy of the theory
appears as an auxiliary mass operator similar to the introduction of the
ground--state Kohn--Sham potential in density functional theory. It is
automatically short--ranged in the same region of Hilbert space which
defines the local Green function. We exploit this property to find good
approximations to the functional. For example, if electronic self--energy is
known to be local in some portion of Hilbert space, a good approximation to
the functional is provided by the corresponding local dynamical mean--field
theory. A simplified implementation of the theory is described based on the
linear muffin--tin orbital method widely used in electronic strucure
calculations. We demonstrate the \ power of the approach on the
long--standing problem of the anomalous volume expansion of metallic
plutonium.
\end{abstract}

\pacs{71.20.-b, 71.27.+a,75.30.-m}
\date{May 2003}
\maketitle

\input{epsf}

\section{Introduction}

Strongly correlated electron systems display remarkably interesting and
puzzling phenomena, such as high--temperature superconductivity, colossal
magnetoresistance, heavy fermion behavior, huge volume expansions and
collapses to name a few. These properties need to be explored with modern
theoretical methods. Unfortunately, the strongly correlated systems are
complex materials with electrons occupying active 3d, 4f or 5f orbitals,
(and sometimes p orbitals as in many organic compounds and in
Bucky--balls--based systems). Here, the excitational spectra over a wide
range of temperatures and frequencies cannot be described in terms of
well--defined quasiparticles. Therefore, the design of computational methods
and algorithms for quantitative description of strongly correlated materials
is a great intellectual challenge, and an enormous amount of work has
addressed this problem in the past \cite%
{ReviewDFT,ReviewGW,ReviewFLEX,ReviewQMC,ReviewExactDiag,ReviewDMRG,ReviewLDA+U,ReviewSIC,ReviewTDDFT,ReviewDMFT,ReviewLDA+DMFT,ReviewTsvelik}%
.

At the heart of the strong--correlation problem is the competition between
localization and delocalization, i.e. between the kinetic energy and the
electron--electron interactions. When the overlap of the electron orbitals
among themselves is large, a wave--like description of the electron is
natural and sufficient. Fermi--liquid theory explains why in a wide range of
energies systems, such as alkali and noble metals, behave as weakly
interacting fermions, i.e. they have a Fermi surface, linear specific heat
and a constant magnetic susceptibility. The one--electron spectra form
quasi--particles and quasi--hole bands and the one--electron spectral
functions show delta--functions like peaks corresponding to the
one--electron excitations. We have powerful quantitative techniques such as
the density functional theory (DFT) in the local density and generalized
gradient approximation (LDA and GGA), for computing ground state properties 
\cite{ReviewDFT}. These techniques can be successfully used as starting
points for perturbative computation of one--electron spectra, for example
using the GW method \cite{ReviewGW}. They have also been successfully used
to compute the strength of the electron--phonon coupling and the resistivity
of simple metals \cite{SavrasovEPI}.

When the electrons are very far apart, a real--space description becomes
valid. A solid is viewed as a regular array of atoms where each element
binds an integer number of electrons. These atoms carry spin and orbital
quantum numbers giving rise to a natural spin and orbital degeneracy.
Transport occurs with the creation of vacancies and doubly occupied sites.
Atomic physics calculations together with perturbation theory around the
atomic limit allow us to derive accurate spin--orbital Hamiltonians. The
one--electron spectrum of the Mott insulators is composed of atomic
excitations which are broaden to form bands that have no single--particle
character. The one--electron Green functions show at least two pole--like
features known as the Hubbard bands \cite{Hubbard1}, and the wave functions
have an atomic--like character, and hence require a many--body description.

The scientific frontier one would like to explore is a category of materials
which falls in between the atomic and band limits. These systems require
both a real space and a momentum space description. To treat these systems
one needs a many--body technique which is able to treat Kohn--Sham bands and
Hubbard bands on the same footing, and which is able to interpolate between
well separated and well overlapping atomic orbitals. The solutions of
many--body equations have to be carried out on the level of the Green
functions which contain necessary information about the total energy and the
spectrum of the solid.

The development of such techniques has a long history in condensed matter
physics. Studies of strongly correlated systems have traditionally focused
on model Hamiltonians using techniques such as diagrammatic methods \cite%
{ReviewFLEX}, Quantum--Monte Carlo simulations \cite{ReviewQMC}, exact
diagonalizations for finite--size clusters \cite{ReviewExactDiag}, density
matrix renormalization group methods\cite{ReviewDMRG} and so on. Model
Hamiltonians are usually written for a given solid--state system based on
physical grounds. In the electronic--structure community, the developments
of LDA+U \cite{ReviewLDA+U} and self--interaction corrected (SIC)\ \cite%
{ReviewSIC} methods , many--body perturbative approaches based on GW and its
extensions \cite{ReviewGW}, as well as time--dependent version of the
density functional theory \cite{ReviewTDDFT} have been carried out. Some of
these techniques are already much more complicated and time--consuming
comparing to the standard LDA based algorithms, and the real exploration of
materials is frequently performed by its simplified versions by utilizing
such, e.g., approximations as plasmon--pole form for the dielectric function 
\cite{PlasmonPole}, omitting self--consistency within GW \cite{ReviewGW} or
assuming locality of the GW self--energy\cite{Zein}.

In general, diagrammatic methods are most accurate if there is a small
parameter in the calculation, say, the ratio of the on--site Coulomb
interaction $U$ to the band width $W$. This does not permit the exploration
of real strongly correlated situations, i.e. when $U/W\sim 1$. Systems near
Mott transition is one of such examples, where strongly renormalized
quasiparticles and atomic--like excitations exist simultaneously. In these
situations, self--consistent methods based on the dynamical mean--field
based theory (DMFT) \cite{ReviewDMFT}, and its cluster generalizations such
as dynamical cluster approximation (DCA) \cite{DCA}, or cellular dynamical
mean field theory (C-DMFT) \cite{CDMFT,Biroli}, are the minimal many body
techniques which have to be employed for exploring real materials.

Thus, a combination of the DMFT based methods with the electronic structure
techniques is promising, because a realistic material--specific description
where the strength of correlation effects is not known \textit{a priori} can
be achieved. This work is in its beginning stages of development but seems
to have a success. The development was started \cite{AnisimovKotliar} by
introducing so--called LDA+DMFT method and applying it to the photoemission
spectrum of La$_{1-x}$Sr$_{x}$TiO$_{3}.$ Near Mott transition, this system
shows a number of features incompatible with the one--electron description 
\cite{LaTiO3exp}. The LDA++ method \cite{LDA++} has been discussed, and the
electronic structure of Fe has been shown to be in better agreement with
experiment than the one based on LDA. The photoemission spectrum near the
Mott transition in V$_{2}$O$_{3}$ has been studied \cite{V2O3}, as well as
issues connected to the finite temperature magnetism of Fe and Ni were
explored \cite{Licht}. LDA\ +DMFT\ was recently generalized to allow
computations of optical properties of strongly correlated materials \cite%
{OpticsDMFT}. Further combinations of the DMFT and GW methods have been
proposed \cite{ReviewTsvelik,Ping,Georges} and a simplified implementation
to Ni has been carried out \cite{Georges}.

Sometimes the LDA+DMFT\ method \cite{ReviewLDA+DMFT} omits full
self--consistency. In this case the approach consists in deriving a model
Hamiltonian with parameters such as the hopping integrals and the Coulomb
interaction matrix elements extracted from an LDA calculation.
Tight--binding fits to the LDA\ energy bands or angular momentum resolved
LDA densities of states for the electrons which are believed to be
correlated are performed. Constrained density functional theory \cite%
{ConstrainedDFT} is used to find the screened on--site Coulomb $U$ and
exchange parameter $J$. This information is used in the downfolded model
Hamiltonian with only active degrees of freedom to explore the consequences
of correlations. Such technique is useful, since it allows us to study real
materials already at the present stage of development. A more ambitious goal
is to build a general method which treats all bands and all electrons on the
same footing, determines both hoppings and interactions internally using a
fully self--consistent procedure, and accesses both energetics and spectra
of correlated materials.

Several ideas to provide a theoretical underpinning to these efforts have
been proposed. The effective action approach to strongly correlated systems
has been used to give realistic DMFT\ an exact functional formulation\cite%
{Chitra1}. Approximations to the exact functional by performing truncations
of the Baym--Kadanoff functional have been discussed \cite{Chitra2}.
Simultaneous treatment of the density and the local Green function in the
functional formulation has been proposed \cite{ReviewTsvelik}. Total energy
calculations using LDA+DMFT have recently appeared in the literature\cite%
{CeMcMahan,PuNature,PuScience,NiOPRL}. DMFT\ corrections have been
calculated and added to the LDA\ total energy in order to explain the
isostructural volume collapse transition in Ce\cite{CeMcMahan}. Fully
self--consistent calculations of charge density, excitation spectrum and
total energy of the $\delta $ phase of metallic Plutonium have been carried
out to address the problem of its anomalous volume expansion \cite{PuNature}%
. The extensions of the method to compute phonon spectra of correlated
systems with the applications to Mott insulators \cite{NiOPRL} and
high--temperature phases of Pu \cite{PuScience} have been also recently
developed.

In this paper we discuss the details of this unified approach which computes
both total energies and spectra of materials with strong correlations and
present our applications for Pu. We utilize the effective action free energy
approach to strongly correlated systems \cite{Chitra1,Chitra2} and write
down the functional of the local Green function. Thus, a spectral density
functional theory (SDFT) is obtained. It can be used to explore strongly
correlated materials from \textit{ab inito} grounds provided useful
approximations exist to the spectral density functional. One of such
approximations is described here, which we refer to as a local dynamical
mean field approximation. It is based on extended \cite{EDMFT} and cluster 
\cite{DCA,CDMFT,Biroli} versions of the dynamical mean--field theory
introduced in connection with the model--Hamiltonian approach \cite%
{ReviewDMFT}.

Implementation of the theory can be carried out on the basis of the
energy--dependent analog for the one--particle wave functions. These are
useful for practical calculations in the same way as Kohn--Sham particles
are used in density functional based calculations. The spectral density
functional theory in its local dynamical mean field approximation, requires
a self--consistent solution of the Dyson equations coupled to the solution
of the Anderson impurity model \cite{AIM} either on a single site \cite%
{ReviewDMFT} or on a cluster \cite{DCA,CDMFT}. Since it is the most
time--consuming part of all DMFT\ algorithms, we are carrying out a
simplified implementation of it based on a slave boson Gtuzwiller \cite%
{Gutz,Ruck,Fleszar} and Hubbard I \cite{Hubbard1,Moments} methods. This is
described in detail in a separate publication\cite{Udo}. We illustrate the
applicability of the method addressing the problem of $\delta -$Pu. Various
aspects of the present work have appeared already \cite%
{ReviewTsvelik,PuNature}.

Our paper is organized as follows. In Section II we describe the spectral
density functional theory and discuss local dynamical mean field
approximation which summarizes the ideas of cluster and extended \cite{EDMFT}
versions of the DMFT. We show that such techniques as LDA+DMFT \cite%
{ReviewLDA+DMFT}, LDA+U \cite{ReviewLDA+U}, and local GW \cite%
{ReviewTsvelik,Zein} methods are naturally seen within the present method.
Section III describes our implementation of the theory based on the
energy--resolved one--particle description \cite{AnisimovKotliar} and
linear--muffin--tin orbital method \cite{OKA-LMTO,TBLMTO,SavrasovLMTO} for
electronic structure calculation. Section IV discusses application of the
method to the volume expansion of Pu. Section V is the conclusion.

\section{Spectral Density Functional Theory}

Here we discuss the basic postulates and approximations of spectral density
functional theory. The central quantity of our formulation is a "local"
Green function $G_{loc}(\mathbf{r},\mathbf{r}^{\prime },z),$ i.e. a part of
the exact electronic Green function which we are interested to compute. This
is by itself arbitrary since we can probe the Green function in a portion of
a certain space such, e.g., as reciprocal space or real space. These are the
most transparent forms where the local Green function can be defined. We can
also probe the Green function in a portion of the Hilbert space. If a
function can be expanded in some basis set $\{\chi _{\xi }\}$%
\begin{equation}
F(\mathbf{r},\mathbf{r}^{\prime },z)=\sum_{\xi \xi ^{\prime }}\chi _{\xi }(%
\mathbf{r})F_{\xi \xi ^{\prime }}(z)\chi _{\xi ^{\prime }}^{\ast }(\mathbf{r}%
^{\prime })  \label{SDFBAS}
\end{equation}%
our interest can, e.g, be associated with diagonal elements of the matrix $%
F_{\xi \xi ^{\prime }}(z)$.

As we see, the locality is a basis set dependent property. Nevertheless, it
is a very useful property because a most economical description of the
function can be achieved. This is true when the basis set which leads to
such description of the function is known. The choice of the appropriate
Hilbert space is therefore crucial if we would like to find an optimal
description of the system with the accuracy proportional to the
computational cost. In spectral density functional theory that has a meaning
of finding good approximations to the functional. Therefore we always rely
on a physical intuition when choosing a particular representation which
should be tailored to a specific physical problem.

\begin{figure}[tbh]
\includegraphics*[height=3.0in]{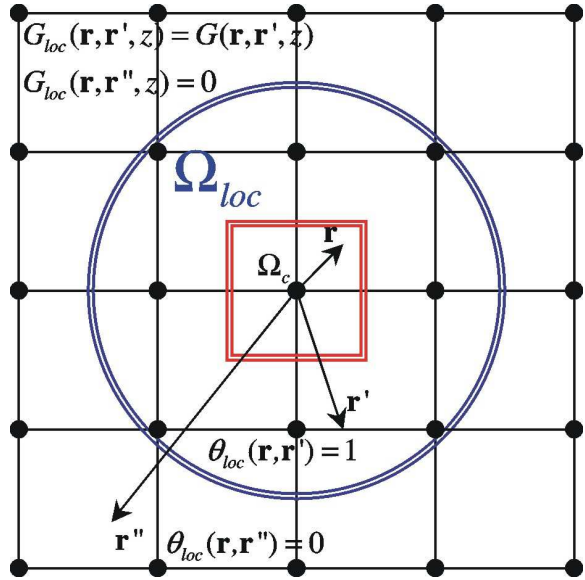}
\caption{Illustration of the area in real space where the local Green
function $G_{loc}$ is defined. Note that $\mathbf{r}$ is restricted by the
unit cell at the origin while $\mathbf{r}^{\prime }$ and $\mathbf{r}^{\prime
\prime }$ travel within the crystal.}
\label{FigGloc}
\end{figure}

At the beginning we formulate spectral density functional theory in
completely real space but keep in mind that such formulation is not unique.
Thus, we are interested in finding a part of the electronic Green function
restricted within a certain cluster area. Due to translational invariance of
the Green function on the original lattice given by primitive translations $%
\{\mathbf{R\}}$, i.e. $G(\mathbf{r+R},\mathbf{r}^{\prime }+\mathbf{R},z)=G(%
\mathbf{r},\mathbf{r}^{\prime },z),$ it is always sufficient to consider $%
\mathbf{r}$ lying within a primitive unit cell $\Omega _{c}$ positioned at $%
\mathbf{R}=0$. Thus, $\mathbf{r}^{\prime }$ travels within some area $\Omega
_{loc}$ centered at $\mathbf{R=0}$. We set the local Green function to be
the exact Green function $G(\mathbf{r},\mathbf{r}^{\prime },z)$ within a
given cluster $\Omega _{loc}$ and zero outside. In other words,%
\begin{equation}
G_{loc}(\mathbf{r},\mathbf{r}^{\prime },z)=G(\mathbf{r},\mathbf{r}^{\prime
},z)\theta _{loc}(\mathbf{r},\mathbf{r}^{\prime })  \label{SDFLOC}
\end{equation}%
where the theta function is a unity when vector $\mathbf{r}\in \Omega _{c},%
\mathbf{r}^{\prime }\in \Omega _{loc}$\ and zero otherwise. It is
schematically illustrated on Fig. \ref{FigGloc}. This construction can be
translationally continued onto entire lattice by enforcing the property $%
\theta _{loc}(\mathbf{r+R},\mathbf{r}^{\prime }+\mathbf{R})=\theta _{loc}(%
\mathbf{r},\mathbf{r}^{\prime }).$

We will now discuss the free energy of a system as a functional of the local
Green function.

\subsection{Functional of Local Green Function}

We consider full many--body Hamiltonian describing the electrons moving in
the periodic ionic potential $V_{ext}(x)=V_{ext}(\mathbf{r})\delta (\tau )$
and interacting among themselves according to the Coulomb law: $%
v_{C}(x-x^{\prime })=e^{2}/|\mathbf{r}-\mathbf{r}^{\prime }|\delta (\tau
-\tau ^{\prime })$ [we use imaginary time--frequency formalism, where $x=(%
\mathbf{r},\tau )$]. This is the formal starting point of our all--electron
first--principles calculation. So, the theory of everything is summarized in
the action $S$:.

\begin{eqnarray}
&&S=\int dx\psi ^{+}(x)[\partial _{\tau }-{\bigtriangledown ^{2}}%
+V_{ext}(x)]\psi (x)  \notag \\
&&+{\frac{{1}}{2}}\int dxdx^{\prime }\psi ^{+}(x)\psi ^{+}(x^{\prime
})v_{C}(x-x^{\prime })\psi (x)\psi (x^{\prime })  \label{SDFactACT}
\end{eqnarray}%
(atomic Rydberg units, $\hbar =1,m_{e}=1/2$, are used throughout). We will
ignore relativistic effects in this action for simplicity but considering
our applications to Pu, these effects will be included later in the
implementation. In addition, the effects of electron--phonon interaction
will not be considered.

We will take the effective action functional approach to describe our
correlated system \cite{Chitra2}. The approach allows to obtain the free
energy of the solid from a functional $\Gamma $ evaluated at its stationary
point. The main question is the choice of the variable of the functional
which is to be extremized. This question is highly non--trivial because the
exact form of the functional is unknown and the usefulness of the approach,
depends on our ability to construct good approximations to it, which in turn
depends on the choice of variables. The Baym--Kadanoff \ (BK) Green function
theory considers exact Green function $G(x,x^{\prime })=-\langle T_{\tau
}\psi (x)\psi ^{+}(x^{\prime })\rangle $ as a variable, i.e. $\Gamma
_{BK}[G].$ Density functional theory considers density $\rho (\mathbf{r})=G(%
\mathbf{r},\mathbf{r},\tau =0)$ of the solid as a variable, i.e. $\Gamma
_{DFT}[\rho ]$. Spectral density functional theory will consider local Green
function $G_{loc}(x,x^{\prime })=$ $G(x,x^{\prime })\theta _{loc}(\mathbf{r},%
\mathbf{r}^{\prime })$ as a variable, i.e. $\Gamma _{SDF}[G_{loc}].$

Notice on the variety of choices we can make, in particular in the
functional $\Gamma _{SDF}[G_{loc}]$ since the definition of the locality is
up to us. The usefulness of a given choice is dictated by the existence of
good approximations to the functional, as, for example, the usefulness of
the DFT is the result of the existence of the\ LDA or GGA, which are
excellent approximations for weakly correlated systems. Here we will argue
that the usefulness of SDFT is the existence of the local dynamical mean
field approximations.

Any of the discussed functionals can be obtained by a Legendre transform of
the effective action. The electronic Green function of a system can be
obtained by probing the system by a source field and monitoring the
response. To obtain $\Gamma _{BK}[G]$ we probe the system with
time--dependent two--variable source field $J(x,x^{\prime })$ or its
imaginary frequency transform $J(\mathbf{r},\mathbf{r}^{\prime },i\omega )$
defined in all space$.$ If we restrict our consideration to saddle point
solutions periodic on the original lattice, we can assume that the field
obeys the periodicity criteria $J(\mathbf{r+R},\mathbf{r}^{\prime }+\mathbf{R%
},i\omega )=J(\mathbf{r},\mathbf{r}^{\prime },i\omega ).$ This restricts the
electronic Green function to be invariant under lattice translations. In
order to obtain a theory based on the density as a physical variable$,$ we
probe the system with a static periodical field $J(\mathbf{r})\delta (\tau ).
$ This delivers \cite{Fukuda,Makov,Fernando} the density functional theory $%
\Gamma _{DFT}[\rho ]$. In order to obtain $\Gamma _{SDF}[G_{loc}]$ we will
probe the system with a local field $J_{loc}(x,x^{\prime })=J_{loc}(\mathbf{r%
},\mathbf{r}^{\prime },\tau -\tau ^{\prime })$ restricted by $\theta _{loc}(%
\mathbf{r},\mathbf{r}^{\prime }).$

Introduction of the time dependent local source $J_{loc}(x,x^{\prime })$
modifies the action of the system (\ref{SDFactACT}) as follows%
\begin{equation}
S^{\prime }[J_{loc}]=S+\int dxdx^{\prime }J_{loc}(x,x^{\prime })\psi
(x^{\prime })\psi ^{+}(x)  \label{SDFactSRC}
\end{equation}%
Due to translational invariance, the integral over $\mathbf{r}$ variable
here is the same for any unit--cell $\Omega _{c}$ and the integral over $%
\mathbf{r}^{\prime }$ should be restricted by the area where $J_{loc}\neq 0$%
, i.e. by the cluster area $\Omega _{loc}.$The average of the operator $\psi
(x)\psi ^{+}(x^{\prime })$ probes the local Green function which is
precisely defined by expression (\ref{SDFLOC}). The partition function $Z,$
or equivalently the free energy of the system $F,$ becomes a functional of
the auxiliary source field%
\begin{equation}
Z[J_{loc}]=\exp (-F[J_{loc}])=\int D[\psi ^{+}\psi ]e^{-S^{\prime }[J_{loc}]}
\label{SDFactPAR}
\end{equation}%
The effective action for the local Green function, i.e., spectral density
functional, is obtained as the Legendre transform of $F$ with respect to the
local Green function $G_{loc}(x,x^{\prime })$, i.e. 
\begin{equation}
\Gamma _{SDF}[G_{loc}]=F[J_{loc}]-\mathrm{Tr}J_{loc}G_{loc}
\label{SDFactSDF}
\end{equation}%
where we use the compact notation $\mathrm{Tr}J_{loc}G_{loc}$ for the
integrals 
\begin{equation}
\mathrm{Tr}J_{loc}G_{loc}=\int dxdx^{\prime }J_{loc}(x,x^{\prime
})G_{loc}(x^{\prime },x)=\sum_{i\omega }\int d\mathbf{r}d\mathbf{r}^{\prime
}J_{loc}(\mathbf{r},\mathbf{r}^{\prime },i\omega )G_{loc}(\mathbf{r}^{\prime
},\mathbf{r},i\omega )  \label{SDFactINT}
\end{equation}%
Using the condition: $J_{loc}=-\delta \Gamma _{SDF}/\delta G_{loc}$ to
eliminate $J_{loc}$ in (\ref{SDFactSDF}) in favor of the local Green
function we finally obtain the functional of the local Green function alone.

The source field sets the degree of locality of the object of interest.
Considering its definition by expanding the cluster till entire solid, we
obtain the Baym--Kadanoff functional which determines the Green function in
all space. Shrinking its definition to a singe point $\mathbf{r}$ and
assuming its frequency (time) independence, i.e. $J(\mathbf{r})\delta (%
\mathbf{r}-\mathbf{r}^{\prime })\delta (\tau -\tau ^{\prime }),$ we obtain
density functional theory. In its extremum, all functionals always reach the
total free energy of the system regardless the choice of the variable. This
situation is similar\ \cite{Makov} to classical thermodynamics where the
thermodynamic potential is either the Helmholtz free energy, or the Gibs
free energy or the entalpy depending on which variables, temperature,
pressure, volume are used. Note also that due to assumed time--dependence of
the source field, away from the extremum the Green function functionals
cannot be interpreted as energies.

Having repeated a formal derivation of the existence \cite{Chitra1} of the
functional $\Gamma _{SDF}[G_{loc}]$ as well as of the functionals $\Gamma
_{BK}[G]$ and $\Gamma _{DFT}[\rho ]$ we now come to the problem of writing
separately various contributions to it. This development parallels the well
known decomposition of the total energy into kinetic energy of a non
interacting system, potential energy, Hartree energy and
exchange--correlation energy. The strategy consists in performing an
expansion of the\ functional in powers of the charge of the electron\cite%
{Chitra1,Fukuda,Fernando,Antoine1,Antoine2}. The lowest order term is the
kinetic part of the action, and the energy associated with the external
potential $V_{ext}$. \ In the Baym Kadanoff Green function theory this term
has the form (\ref{SDFactACT}):%
\begin{equation}
K_{BK}[G]=\mathrm{Tr}\ln G-\mathrm{Tr}[G_{0}^{-1}-G^{-1}]G  \label{SDFactKBK}
\end{equation}%
The $G_{0}(\mathbf{r},\mathbf{r}^{\prime },i\omega )$ is the
non--interacting Green function, which is given by%
\begin{eqnarray}
G_{0}^{-1}(\mathbf{r},\mathbf{r}^{\prime },i\omega ) &=&\delta (\mathbf{r}-%
\mathbf{r}^{\prime })[i\omega +\mu +\nabla ^{2}-V_{ext}(\mathbf{r})]
\label{SDFactG01} \\
\delta (\mathbf{r}-\mathbf{r}^{\prime }) &=&\int d\mathbf{r}^{\prime \prime
}G_{0}^{-1}(\mathbf{r},\mathbf{r}^{\prime \prime },i\omega )G_{0}(\mathbf{r}%
^{\prime \prime },\mathbf{r}^{\prime },i\omega )  \label{SDFactG11}
\end{eqnarray}%
where $\mu $ is a chemical potential. Note that since finite temperature
formulation is adopted we did not obtain simply $K_{BK}[G]=\mathrm{Tr(}%
-\nabla ^{2}+V_{ext})G$ but also have got all entropy based contributions.

Let us now turn to the density functional theory. In principle, it does not
have a closed formula to describe fully interacting kinetic energy as the
density functional. However, it solves this problem by introducing a
non--interacting part of the kinetic energy. It is described by its own
Green function $G_{KS}(\mathbf{r},\mathbf{r}^{\prime },i\omega ),$ which is
related to the Kohn--Sham (KS) representation. An auxiliary set of
non--interacting particles is introduced which is used to mimic the density
of the system. These particles move in some effective one--particle
Kohn--Sham potential $V_{eff}(\mathbf{r})=V_{ext}(\mathbf{r})+V_{int}(%
\mathbf{r})$. This potential is chosen merely to reproduce the density and
does not have any other physical meaning at this point. The Kohn--Sham Green
function is defined in the entire space by the relation $G_{KS}^{-1}(\mathbf{%
r},\mathbf{r}^{\prime },i\omega )=G_{0}^{-1}(\mathbf{r},\mathbf{r}^{\prime
},i\omega )-V_{int}(\mathbf{r})\delta (\mathbf{r}-\mathbf{r}^{\prime })$,
where $V_{int}(\mathbf{r})$ is adjusted so that the density of the system $%
\rho (\mathbf{r})$ can be found from $G_{KS}(\mathbf{r},\mathbf{r}^{\prime
},i\omega )$. Since the exact Green function $G$ and the local Green
function $G_{loc}$ can be also used to find the density, we can write a
general relationship:%
\begin{equation}
\rho (\mathbf{r})=T\sum_{i\omega }G_{KS}(\mathbf{r},\mathbf{r},i\omega
)e^{i\omega 0+}=T\sum_{i\omega }G(\mathbf{r},\mathbf{r},i\omega )e^{i\omega
0+}=T\sum_{i\omega }G_{loc}(\mathbf{r},\mathbf{r},i\omega )e^{i\omega 0+}
\label{SDFactRHO}
\end{equation}%
where the sum over $i\omega $ assumes the summation on the Matsubara axis at
given temperature $T$. With the introduction of $G_{KS}$ the
non--interacting kinetic portion of the action plus the energy related to $%
V_{ext}$ can be written in complete analogy with (\ref{SDFactKBK}) as follows

\begin{equation}
K_{DFT}[G_{KS}]=\mathrm{Tr}\ln G_{KS}-\mathrm{Tr}%
[G_{0}^{-1}-G_{KS}^{-1}]G_{KS}  \label{SDFactKDF}
\end{equation}

\begin{figure}[tbh]
\includegraphics*[height=3.0in]{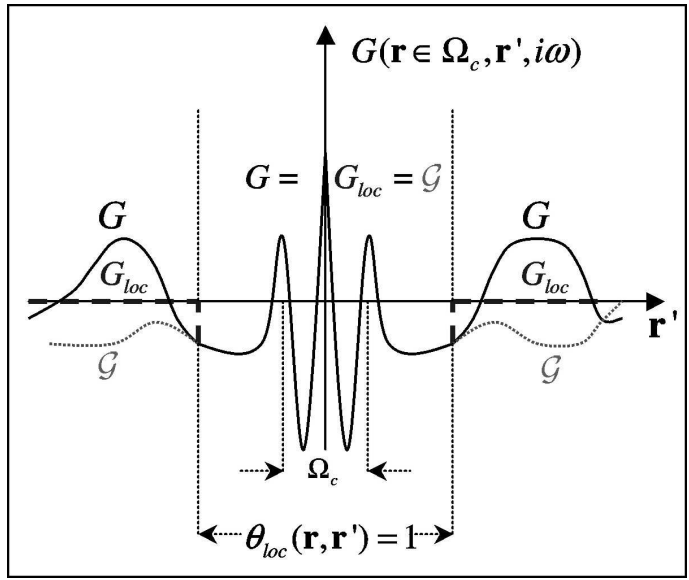}
\caption{Relationship between various Green functions in spectral density
functional theory: exact Green function $G$, local Green function $G_{loc}$
and auxiliary Green function $\mathcal{G}$ are the same in a certain region
of space of our interest. They are all different outside this area, where
the local Green function is zero by definiton.}
\label{FigGG0GC}
\end{figure}

In order to describe the different contributions to the thermodynamical
potential in the spectral density functional theory, we introduce a notion
of the energy--dependent analog of Kohn--Sham representation. These
auxiliary particles are interacting so that they will describe not only the
density but also a local part of the Green function of the system, and will
feel a frequency dependent potential. The latter is a field described by
some effective mass operator $\mathcal{M}_{eff}(\mathbf{r},\mathbf{r}%
^{\prime },i\omega )=V_{ext}(\mathbf{r})\delta (\mathbf{r}-\mathbf{r}%
^{\prime })+\mathcal{M}_{int}(\mathbf{r},\mathbf{r}^{\prime },i\omega ).$ We
now introduce an auxiliary Green function $\mathcal{G}(\mathbf{r},\mathbf{r}%
^{\prime },i\omega )$ connected to our new "interacting Kohn--Sham"
particles so that it is defined in the entire space by the relationship $%
\mathcal{G}^{-1}(\mathbf{r},\mathbf{r}^{\prime },i\omega )=G_{0}^{-1}(%
\mathbf{r},\mathbf{r}^{\prime },i\omega )-\mathcal{M}_{int}(\mathbf{r},%
\mathbf{r}^{\prime },i\omega )$. Thus, $\mathcal{M}_{int}(\mathbf{r},\mathbf{%
r}^{\prime },i\omega )$ is\ a function which has the same range as the
source that we introduce: it is adjusted until the auxiliary $\mathcal{G}(%
\mathbf{r},\mathbf{r}^{\prime },i\omega )$ coincides with the local Green
function inside the area restricted by $\theta _{loc}(\mathbf{r},\mathbf{r}%
^{\prime }),$ i.e 
\begin{equation}
G_{loc}(\mathbf{r},\mathbf{r}^{\prime },i\omega )=\mathcal{G}(\mathbf{r},%
\mathbf{r}^{\prime },i\omega )\theta _{loc}(\mathbf{r},\mathbf{r}^{\prime })
\label{SDFactGSK}
\end{equation}%
We illustrate the relationship between all introduced Green functions in
Fig. \ref{FigGG0GC}. Note that $\mathcal{G}(\mathbf{r},\mathbf{r}^{\prime
},i\omega )$ also delivers the exact density of the system. With the help of 
$\mathcal{G}$ the kinetic term in the spectral density functional theory can
be represented as follows%
\begin{equation}
K_{SDF}[\mathcal{G]}=Tr\ln \mathcal{G}-Tr[G_{0}^{-1}-\mathcal{G}^{-1}]%
\mathcal{G}  \label{SDFactKSD}
\end{equation}

Since $G_{KS}$ is a functional of $\rho $, DFT\ considers the density
functional as the functional of Kohn--Sham wave functions, i.e. as $\Gamma
_{DFT}[G_{KS}].$ Similarly, since $\mathcal{G}$ is a functional of $G_{loc}$%
, it is very useful to view the spectral density functional $\Gamma _{SDF}$
as a functional of $\mathcal{G}$:%
\begin{equation}
\Gamma _{SDF}[\mathcal{G}]=\mathrm{Tr}\ln \mathcal{G}-\mathrm{Tr}[G_{0}^{-1}-%
\mathcal{G}^{-1}]\mathcal{G}+\Phi _{SDF}[G_{loc}]  \label{SDFactSD2}
\end{equation}%
where the unknown interaction part of the free energy $\Phi _{SDF}[G_{loc}]$
is the functional of $G_{loc}.$ If the Hartree term is explicitly extracted,
this functional can be represented as%
\begin{equation}
\Phi _{SDF}[G_{loc}]=E_{H}[\rho ]+\Phi _{SDF}^{xc}[G_{loc}]
\label{SDFactFXC}
\end{equation}%
where $E_{H}[\rho ]$ is the Hartree energy depending only on the density of
the system, and where $\Phi _{SDF}^{xc}[G_{loc}]$ is the
exchange--correlation part of the free energy. Notice that the density of
the system can be obtained via $G_{loc}$ or $\mathcal{G},$ therefore the
Hartree term can be also viewed as a functional of $G_{loc}$ or $\mathcal{G}%
. $ Notice also, that since the kinetic energies (\ref{SDFactKBK}), (\ref%
{SDFactKDF}), (\ref{SDFactKSD}) are defined differently in all theories, the
interaction energies $\Phi _{SDF}[G_{loc}],$ $\Phi _{BK}[G],$ $\Phi
_{DFT}[\rho ]$ are also different.

The stationarity of the spectral density functional can be examined with
respect to $\mathcal{G}$ 
\begin{equation}
\frac{\delta \Gamma _{SDF}}{\delta \mathcal{G}(\mathbf{r},\mathbf{r}^{\prime
},i\omega )}=0  \label{SDFactSD0}
\end{equation}%
similar to the stationarity conditions for $\Gamma _{BK}[G]$ and $\Gamma
_{DFT}[G_{KS}]$ 
\begin{eqnarray}
\frac{\delta \Gamma _{BK}}{\delta G(\mathbf{r},\mathbf{r}^{\prime },i\omega )%
} &=&0  \label{SDFactGB0} \\
\frac{\delta \Gamma _{DFT}}{\delta G_{KS}(\mathbf{r},\mathbf{r}^{\prime
},i\omega )} &=&0  \label{SDFactKS0}
\end{eqnarray}%
This leads to the equations for the corresponding Green functions in all
theories:%
\begin{equation}
\mathcal{G}^{-1}(\mathbf{r},\mathbf{r}^{\prime },i\omega )=G_{0}^{-1}(%
\mathbf{r},\mathbf{r}^{\prime },i\omega )-\mathcal{M}_{int}(\mathbf{r},%
\mathbf{r}^{\prime },i\omega )  \label{SDFactC01}
\end{equation}%
as well as%
\begin{eqnarray}
G^{-1}(\mathbf{r},\mathbf{r}^{\prime },i\omega ) &=&G_{0}^{-1}(\mathbf{r},%
\mathbf{r}^{\prime },i\omega )-\Sigma _{int}(\mathbf{r},\mathbf{r}^{\prime
},i\omega )  \label{SDFactBK1} \\
G_{KS}^{-1}(\mathbf{r},\mathbf{r}^{\prime },i\omega ) &=&G_{0}^{-1}(\mathbf{r%
},\mathbf{r}^{\prime },i\omega )-V_{int}(\mathbf{r})\delta (\mathbf{r}-%
\mathbf{r}^{\prime })  \label{SDFactKS1}
\end{eqnarray}%
By using (\ref{SDFactG01}) for $G_{0}^{-1}$ and by multiplying both parts by
the corresponding Green functions we obtain familiar Dyson equations%
\begin{equation}
\lbrack -\nabla ^{2}+V_{ext}(\mathbf{r})-i\omega -\mu ]\mathcal{G}(\mathbf{r}%
,\mathbf{r}^{\prime },i\omega )+\int d\mathbf{r}^{\prime \prime }\mathcal{M}%
_{int}(\mathbf{r},\mathbf{r}^{\prime \prime },i\omega )\mathcal{G}(\mathbf{r}%
^{\prime \prime },\mathbf{r}^{\prime },i\omega )=\delta (\mathbf{r}-\mathbf{r%
}^{\prime })  \label{SDFactDEC}
\end{equation}%
and%
\begin{eqnarray}
\lbrack -\nabla ^{2}+V_{ext}(\mathbf{r})-i\omega -\mu ]G(\mathbf{r},\mathbf{r%
}^{\prime },i\omega )+\int d\mathbf{r}^{\prime \prime }\Sigma _{int}(\mathbf{%
r},\mathbf{r}^{\prime \prime },i\omega )G(\mathbf{r}^{\prime \prime },%
\mathbf{r}^{\prime },i\omega ) &=&\delta (\mathbf{r}-\mathbf{r}^{\prime })
\label{SDFactDEF} \\
\lbrack -\nabla ^{2}+V_{ext}(\mathbf{r})-i\omega -\mu ]G_{KS}(\mathbf{r},%
\mathbf{r}^{\prime },i\omega )+V_{int}(\mathbf{r})G_{KS}(\mathbf{r}^{\prime
\prime },\mathbf{r}^{\prime },i\omega ) &=&\delta (\mathbf{r}-\mathbf{r}%
^{\prime })  \label{SDFactDES}
\end{eqnarray}%
The stationarity condition brings the definition of the auxiliary mass
operator $\mathcal{M}_{int}(\mathbf{r},\mathbf{r}^{\prime },i\omega )$ which
is the variational derivative of the interaction free energy with respect to
the local Green function:%
\begin{equation}
\mathcal{M}_{int}(\mathbf{r},\mathbf{r}^{\prime },i\omega )=\frac{\delta
\Phi _{SDF}[G_{loc}]}{\delta \mathcal{G}(\mathbf{r}^{\prime },\mathbf{r}%
,i\omega )}=\frac{\delta \Phi _{SDF}[G_{loc}]}{\delta G_{loc}(\mathbf{r}%
^{\prime },\mathbf{r},i\omega )}\theta _{loc}(\mathbf{r},\mathbf{r}^{\prime
})  \label{SDFactSIG}
\end{equation}%
It plays the role of the effective self--energy which is short--ranged
(local) in the space. The corresponding expressions hold for the interaction
parts of the exact self--energy of the electron $\Sigma _{int}(\mathbf{r},%
\mathbf{r}^{\prime },i\omega )$ and for the interaction part of the
Kohn--Sham potential $V_{int}(\mathbf{r}).$%
\begin{eqnarray}
\Sigma _{int}(\mathbf{r},\mathbf{r}^{\prime },i\omega ) &=&\frac{\delta \Phi
_{BK}[G]}{\delta G(\mathbf{r}^{\prime },\mathbf{r},i\omega )}
\label{SDFactSBK} \\
V_{int}(\mathbf{r})\delta (\mathbf{r}-\mathbf{r}^{\prime }) &=&\frac{\delta
\Phi _{DFT}[\rho ]}{\delta G_{KS}(\mathbf{r}^{\prime },\mathbf{r},i\omega )}=%
\frac{\delta \Phi _{DFT}[\rho ]}{\delta \rho (\mathbf{r})}\delta (\mathbf{r}-%
\mathbf{r}^{\prime })  \label{SDFactSKS}
\end{eqnarray}%
If the external potential is added to these quantities we obtain total
effective self--energies/potentials of the SDF, BK and DF theories: $%
\mathcal{M}_{eff}(\mathbf{r},\mathbf{r}^{\prime },i\omega ),$ $\Sigma _{eff}(%
\mathbf{r},\mathbf{r}^{\prime },i\omega ),$ $V_{eff}(\mathbf{r})$
respectively. If the Hartree potential $V_{H}(\mathbf{r})$ is separated we
obtain the exchange--correlation parts: $\mathcal{M}_{xc}(\mathbf{r},\mathbf{%
r}^{\prime },i\omega ),$ $\Sigma _{xc}(\mathbf{r},\mathbf{r}^{\prime
},i\omega ),$ $V_{xc}(\mathbf{r}).$

Note that strictly speaking the substitution of variables, $G_{KS}$ vs. $%
\rho ,$ in the density functional as well as the substitution of variables, $%
\mathcal{G}$ vs. $G_{loc}$, in the spectral density functional is only
possible under the assumption of the so--called $V$--representability (or $M$%
--representability), i.e. the existence of such effective potential (mass
operator) which can be used to construct the exact density (local Green
function) of the system via the non--interacting Kohn--Sham particles of the
DFT or its energy--dependent generalization in SDFT.

Note also that the effective mass--operator of spectral density functional
theory is local by construction, i.e. it is non--zero only within the
cluster area $\Omega _{loc}$ restricted by $\theta _{loc}(\mathbf{r},\mathbf{%
r}^{\prime }).$ It is an auxiliary object which cannot be identified with
the exact self--energy of the electron $\Sigma _{eff}(\mathbf{r},\mathbf{r}%
^{\prime },i\omega ).$ This is similar to the observation that the
Kohn--Sham potential of the DFT cannot be associated with the exact
self--energy as well. Nevertheless, the SDFT always delivers local Green
function and the total free energy exactly (at least in principle) as long
as the exact functional is used. In the limit when the exact self--energy of
the electron is indeed localized within $\Omega _{loc}$, the SDFT becomes
the Baym--Kadanoff functional which delivers the full Green function of the
system, i.e. we can immediately identify $\mathcal{M}_{eff}(\mathbf{r},%
\mathbf{r}^{\prime },i\omega )$ with $\Sigma _{eff}(\mathbf{r},\mathbf{r}%
^{\prime },i\omega )$ and the poles of $\mathcal{G}(\mathbf{r},\mathbf{r}%
^{\prime },i\omega )$ with exact poles of $G(\mathbf{r},\mathbf{r}^{\prime
},i\omega )$ where the information about both k and energy dependence as
well as life time of the quasiparticles is contained$.$ We thus see that, at
least formally, increasing the size of $\Omega _{loc}$ \ in the SDF\ theory
leads to a complete description of the many--body system, the situation
quite different from the DFT which misses such scaling.

From a conceptual point of view, the spectral density functional approach
constitutes a radical departure from the DFT\ philosophy. The saddle--point
equation (\ref{SDFactDEC}) is the equation for a continuous distribution of
spectral weight and the obtained local spectral function $G_{loc}$ can now
be identified with the observable local (roughly speaking, k--integrated)
one--electron spectrum. This is very different from the Kohn--Sham
quasiparticles which are the poles of $G_{KS}$ not identifiable rigorously
with any one--electron excitations. While the SDFT approach is
computationally more demanding than DFT, it is formulated in terms of
observables and gives more information than DFT.

On one side, spectral density functional can be viewed as approximation or
truncation of the full Baym Kadanoff theory where $\Phi _{BK}[G]$ is
approximated by $\Phi _{SDF}[G_{loc}]$ \ by restricting \ $G$ to $G_{loc}$%
\cite{Chitra2} and the kinetic functionals $K_{BK\text{ }}[G]$ and $K_{SDF}[%
\mathcal{G}]$ are thought to be the same. Such restriction will
automatically generate a short--ranged self--energy in the theory. This is
similar to the interpretation of DFT as approximation $\Phi _{BK}[G]=\Phi
_{DFT}[\rho ],K_{BK\text{ }}[G]$=$K_{SDF}[G_{KS}]$ which would generate the
DFT potential as the self--energy. However, SDFT can be thought as a
separate theory whose manifestly local self--energy is an auxiliary operator
introduced to reproduce the local part of the Green function of the system,
exactly like the Kohn--Sham ground state potential is an auxiliary operator
introduced to reproduce the density of the electrons in DFT.

Spectral density functional theory contains the exchange--correlation
functional $\Phi _{SDF}[G_{loc}]$. An explicit expression for it involving a
coupling constant $\lambda =e^{2}$ \ integration can be obtained in complete
analogy with the Harris--Jones formula\cite{Harris} of density functional
theory\cite{Antoine2}. One considers $\Gamma _{SDF}[\mathcal{G},\lambda ]$
at an arbitrary interaction $\lambda $ and expresses 
\begin{equation}
\Gamma _{SDF}[\mathcal{G},e^{2}]=\Gamma _{SDF}[\mathcal{G}%
,0]+\int_{0}^{e^{2}}d\lambda \frac{\partial \Gamma _{SDF}[\mathcal{G}%
,\lambda ]}{\partial \lambda }  \label{SDFactGU1}
\end{equation}

Here the first term is simply the kinetic part $K_{SDF}[\mathcal{G]}$ as
given by (\ref{SDFactKSD}) which does not depend on $\lambda $. The second
part is thus the unknown functional $\Phi _{SDF}[G_{loc}].$ The derivative
with respect to the coupling constant in (\ref{SDFactACT}) \ is given by the
average $\langle \psi ^{+}(x)\psi ^{+}(x^{\prime })\psi (x)\psi (x^{\prime
})\rangle $ $=\Pi _{\lambda }(x,x^{\prime },i\omega )+$ $\langle \psi
^{+}(x)\psi (x)\rangle \langle \psi ^{+}(x^{\prime })\psi (x^{\prime
})\rangle $ where $\Pi _{\lambda }(x,x^{\prime })$ is the density--density
correlation function at a given interaction strength $\lambda $ computed in
the presence of a source which is $\lambda $ dependent \ and chosen so that
the local Greens function of the system is $\mathcal{G}$.\ Since $\langle
\psi ^{+}(x)\psi (x)\rangle =\rho (\mathbf{r})\delta (\tau ),$ we can obtain
:%
\begin{equation}
\Phi _{SDF}[G_{loc}]=E_{H}[\rho ]+\sum_{i\omega }\int_{0}^{e^{2}}d\lambda 
\frac{\Pi _{\lambda }(\mathbf{r},\mathbf{r}^{\prime },i\omega )}{|\mathbf{r}-%
\mathbf{r}^{\prime }|}  \label{SDFactGU2}
\end{equation}%
\bigskip

Establishing the diagrammatic rules for the functional $\Phi _{SDF}[G_{loc}]$
while possible \cite{Chitra1}, is not as simple as for the functional $\Phi
_{BK}[G].$ The latter is formally represented as a sum of two--particle
diagrams constructed with $G$ and $v_{C}.$ It is well known that instead of
expanding $\Phi _{BK}[G]$ in powers of the bare interaction $v_{C}$ and $G,$
the functional form can be obtained by introducing the dynamically screened
Coulomb interaction $W(\mathbf{r},\mathbf{r}^{\prime },i\omega )$ as a
variable. In the effective action formalism \cite{Chitra2} this was done by
introducing an auxiliary Bose variable coupled to the density, which
transforms the original problem into a problem of electrons interacting with
the Bose field. $W$ is the connected correlation function of the Bose field.

Our effective action is now a functional of $G$, $W$ and of the expectation
value of the Bose field. Since the latter couples linearly to the density it
can be eliminated exactly, a step which generates the Hartree term. After
this elimination, the functional takes the form 
\begin{equation}
\Gamma _{BK}[G,W]=\mathrm{Tr}\ln G-\mathrm{Tr}[G_{0}^{-1}-G^{-1}]G+\Phi
_{BK}[G,W]  \label{SDFactGBK}
\end{equation}

\begin{equation}
\Phi _{BK}[G,W]=E_{H}[\rho ]-\frac{1}{2}\mathrm{Tr}\ln W+\frac{1}{2}\mathrm{%
Tr}[v_{C}^{-1}-W^{-1}]W+\Psi _{BK}[G,W]  \label{SDFactPSI}
\end{equation}%
The entire theory is viewed as the functional of both $G$ and $W.$ Here, $%
\Psi _{BK}[G,W]$ is the sum of all two--particle diagrams constructed with $%
G $ and $W$ with the exclusion of the Hartree term, which is evaluated with
the bare Coulomb interaction. An additional stationarity condition $\delta
\Gamma _{BK}/\delta W=0$ leads to the equation for the screened Coulomb
interaction $W$%
\begin{equation}
W^{-1}(\mathbf{r},\mathbf{r}^{\prime },i\omega )=v_{C}^{-1}(\mathbf{r}-%
\mathbf{r}^{\prime })-\Pi (\mathbf{r},\mathbf{r}^{\prime },i\omega )
\label{SDFactDEW}
\end{equation}%
where the function $\Pi (\mathbf{r},\mathbf{r}^{\prime },i\omega )=-2\delta
\Psi _{BK}/\delta W(\mathbf{r},\mathbf{r}^{\prime },i\omega )$ is the exact
interacting susceptibility of the system, which is already discussed in
connection with representation (\ref{SDFactGU2}).

A similar theory is developed for the local quantities \cite{Chitra2}, and
this generalization represents the ideas of extended dynamical mean field
theory \cite{EDMFT}, now viewed as an exact theory. Namely, one constructs
an exact functional of the local Greens function and the local\ correlator
of the Bose field coupled to the density which can be identified with the
local part of the dynamically screened interaction. The real--space
definition of it is the following%
\begin{equation}
W_{loc}(\mathbf{r},\mathbf{r}^{\prime },i\omega )=W(\mathbf{r},\mathbf{r}%
^{\prime },i\omega )\theta _{loc}(\mathbf{r},\mathbf{r}^{\prime })
\label{SDFactWLC}
\end{equation}%
which is non--zero within a given cluster $\Omega _{loc}$. Note that
formally this cluster can be different from the one considered to define the
local Green function (\ref{SDFLOC}) but we will not distinguish between them
for simplicity. An auxiliary interaction $\mathcal{W}(\mathbf{r},\mathbf{r}%
^{\prime },i\omega )$ is introduced which is the same as the local part of
the exact interaction within non--zero area of $\theta _{loc}(\mathbf{r},%
\mathbf{r}^{\prime })$%
\begin{equation}
W_{loc}(\mathbf{r},\mathbf{r}^{\prime },i\omega )=\mathcal{W}(\mathbf{r},%
\mathbf{r}^{\prime },i\omega )\theta _{loc}(\mathbf{r},\mathbf{r}^{\prime })
\label{SDFactWLO}
\end{equation}%
The interaction part of the spectral density functional is represented in
the form similar to (\ref{SDFactPSI})%
\begin{equation}
\Phi _{SDF}[G_{loc},W_{loc}]=E_{H}[\rho ]-\frac{1}{2}\mathrm{Tr}\ln \mathcal{%
W}+\frac{1}{2}\mathrm{Tr}[v_{C}^{-1}-\mathcal{W}^{-1}]\mathcal{W}+\Psi
_{SDF}[G_{loc},W_{loc}]  \label{SDFactFP1}
\end{equation}%
and the spectral density functional is viewed as a functional $\Gamma
_{SDF}[G_{loc},W_{loc}]$\ or alternatively as a functional $\Gamma _{SDF}[%
\mathcal{G},\mathcal{W}]$. $\Psi _{SDF}[G_{loc},W_{loc}]$ is formally \emph{%
not} a sum of two--particle diagrams constructed with $G_{loc}\ $and $%
W_{loc} $, but in principle a more complicated diagrammatic expression can
be derived. Alternatively, a more explicit expression involving a coupling
constant integration can be given. Examining stationarity $\delta \Gamma
_{SDF}/\delta \mathcal{W}=0$ yields a saddle--point equation for $\mathcal{W}%
(\mathbf{r},\mathbf{r}^{\prime },i\omega )$%
\begin{equation}
\mathcal{W}^{-1}(\mathbf{r},\mathbf{r}^{\prime },i\omega )=v_{C}^{-1}(%
\mathbf{r}-\mathbf{r}^{\prime })-\mathcal{P}(\mathbf{r},\mathbf{r}^{\prime
},i\omega )  \label{SDFactWM1}
\end{equation}%
\bigskip where the effective susceptibility of the system is the variational
derivative%
\begin{equation}
\mathcal{P}(\mathbf{r},\mathbf{r}^{\prime },i\omega )=\frac{-2\delta \Psi
_{SDF}}{\delta \mathcal{W}(\mathbf{r}^{\prime },\mathbf{r},i\omega )}=\frac{%
-2\delta \Psi _{SDF}}{\delta W_{loc}(\mathbf{r}^{\prime },\mathbf{r},i\omega
)}\theta _{loc}(\mathbf{r},\mathbf{r}^{\prime })  \label{SDFactCHW}
\end{equation}%
Notice again a set of parallel observations for $\mathcal{P}$ as for $%
\mathcal{M}_{eff}$, Eq. (\ref{SDFactSIG}). The effective susceptibility of
spectral density functional theory is local by construction, i.e. it is
non--zero only within the cluster restricted by $\theta _{loc}(\mathbf{r},%
\mathbf{r}^{\prime }).$ Formally, it is an auxiliary object and cannot be
identified with the exact susceptibility of the electronic system $\Pi (%
\mathbf{r},\mathbf{r}^{\prime },i\omega ).$ However, if the exact
susceptibility $\Pi (\mathbf{r},\mathbf{r}^{\prime },i\omega )$ is
sufficiently localized, this identification becomes possible. If cluster $%
\Omega _{loc}$ includes physical area of localization, we can immediately
identify $\mathcal{P}(\mathbf{r},\mathbf{r}^{\prime },i\omega )$ with $\Pi (%
\mathbf{r},\mathbf{r}^{\prime },i\omega )$ and $\mathcal{W}(\mathbf{r},%
\mathbf{r}^{\prime },i\omega )$ with $W(\mathbf{r},\mathbf{r}^{\prime
},i\omega )$ in all space. However, both $\mathcal{W}$ and $W$ are always
the same within $\Omega _{loc}$ regardless its size, as it is seen from (\ref%
{SDFactWLC}) and (\ref{SDFactWLO}).

At the stationarity point, $\Gamma _{SDF}[\mathcal{G},\mathcal{W}]$ is the
free energy $F$ of the system. If one inserts (\ref{SDFactC01}) into (\ref%
{SDFactKSD}) and (\ref{SDFactWM1}) into (\ref{SDFactFP1}) we obtain the
formula:%
\begin{equation}
F=\mathrm{Tr}\ln \mathcal{G}-\mathrm{Tr}\mathcal{M}_{eff}\mathcal{G}+\mathrm{%
Tr}V_{ext}\mathcal{G+}E_{H}-\frac{1}{2}\mathrm{Tr}\ln \mathcal{W}+\frac{1}{2}%
\mathrm{Tr}\mathcal{PW}+\Psi _{SDF}  \label{SDFactFRE}
\end{equation}%
Similar formulae hold for the Baym--Kadanoff and density functional theories%
\begin{eqnarray}
F &=&\mathrm{Tr}\ln G-\mathrm{Tr}\Sigma _{eff}G+\mathrm{Tr}V_{ext}G+E_{H}-%
\frac{1}{2}\mathrm{Tr}\ln W+\frac{1}{2}\mathrm{Tr}\Pi W+\Psi _{BK}
\label{SDFactBKF} \\
F &=&\mathrm{Tr}\ln G_{KS}-\mathrm{Tr}V_{eff}G_{KS}+\mathrm{Tr}%
V_{ext}G_{KS}+\Phi _{DFT}  \label{SDFactDFF}
\end{eqnarray}%
where the first two terms in all expressions (\ref{SDFactFRE}), (\ref%
{SDFactBKF}), (\ref{SDFactDFF}) are interpreted as corresponding kinetic
energies, the third term is the energy related to the external potential $%
V_{ext}$ which is in fact $\mathrm{Tr}V_{ext}\rho $ in all cases. The other
terms represent the interaction parts of the free energy. Note that all
entropy originated contributions are included in both kinetic and
interaction parts. If temperature goes to zero, the entropy part disappears
and the total energy formulae will be recovered. For example, in spectral
density functional theory we obtain:%
\begin{equation}
E=-\mathrm{Tr}\nabla ^{2}\mathcal{G}+\mathrm{Tr}V_{ext}\rho +E_{H}+\Phi _{xc}
\label{SDFactTOT}
\end{equation}
We will also discuss this limit later in more details in Section III.

The SDFT approach is so far not very useful since a tractable expression for
the functional form of $\Phi _{SDF}[G_{loc}]$ or $\Psi
_{SDF}[G_{loc},W_{loc}]$ has not been given yet. This is quite similar to
the unknown exchange--correlation functional of the DFT. As we have learned
from the developments of the dynamical mean--field methods, a very useful
approximation exists to access these functionals. This is based on a full
many--body solution of a finite--size cluster problem treated as an impurity
embedded into a bath subjected to a self--consistency condition. Such local
dynamical mean field theory will be discussed below.

\subsection{Local Dynamical Mean Field Approximation}

The spectral density functional theory, where an exact functional of \
certain local quantities is constructed in the spirit of Ref. %
\onlinecite{Chitra1} uses effective self--energies and susceptibilities
which are local by construction. This property can be exploited to find good
approximations to the interaction energy functional. For example, if it is 
\textit{a priori} known that the real electronic self--energy is local in a
certain portion of the Hilbert space, a good approximation is the
corresponding local dynamical mean field theory obtained for example by a
restriction or truncation of the full Baym--Kadanoff functional or its
generalization to use $W$ and $G$ as natural variables, to local quantities
in the spirit of Ref. \onlinecite{Chitra2}.

The local DMFT approximates the functional $\Phi _{SDF}[G_{loc}]$ (or $\Psi
_{SDF}[G_{loc},W_{loc}])$ by the sum of all two--particle diagrams evaluated
with $G_{loc}$ and the bare Coulomb interaction $v_{C}$ (or screened local
interaction $W_{loc}).$ In other words, the functional dependence of the
interaction part $\Phi _{BK}[G]$ in the Baym--Kadanoff functional for which
the diagrammatic rules exist is now restricted by $G_{loc}$ and is used as
an approximation to $\Phi _{SDF}[G_{loc}]$, i.e. $\Phi _{SDF}[G_{loc}]=\Phi
_{BK}[G_{loc}].$ Obviously that the variational derivative of such
restricted functional will generate the self--energy confined in the same
area as the local Green function itself.

Remarkably the summation over all local diagrams can be performed exactly
via introduction of an auxiliary quantum impurity model subjected to a
self--consistency condition \cite{Georges92,ReviewDMFT}. If this impurity is
considered as a cluster $C$, the cellular DMFT (C--DMFT) can be used which
breaks the translational invariance of the lattice to obtain accurate
estimates of the self energies. The C--DMFT approximation, can also be
motivated using the cavity construction. The solid should be separated onto
large cells which circumscribe the areas $\Omega _{loc}$. Considering the
effective action $S,$ Eq. (\ref{SDFactACT}), the integration volume is
separated onto the cellular area $\Omega _{C}$ and the rest bath area $%
\Omega -\Omega _{C}=\Omega _{bath}.$ The action is now represented as the
action of the cluster cell, $\Omega _{C}$ plus the action of the bath,$\
\Omega _{bath},\ $plus the interaction between those two. We are interested
in the local effective action $S_{C}$ of the cluster degrees of freedom
only, which is obtained conceptually by integrating out the bath in the
functional integral:

\begin{equation}
\frac{1}{Z_{C}}\exp [-S_{C}]=\frac{1}{Z}\int_{\Omega _{bath}}d\mathbf{r}d%
\mathbf{r}^{\prime }\exp [-S]  \label{SDFlsdINT}
\end{equation}%
where $Z_{C}$ and $Z$\ are the corresponding partition functions.\ This
integration \ is carried out approximately, keeping only a charge--charge
interaction as quartic terms and neglecting all the higher order terms
generated in this process to arrive to a cavity action of the form \cite%
{EDMFT,Chitra2,CDMFT,Ping}: 
\begin{eqnarray}
&&S_{C}=-\int dx\psi ^{+}(x)\mathcal{G}_{0}^{-1}(x,x^{\prime })\psi
(x^{\prime })  \notag \\
&&+{\frac{{1}}{2}}\int dxdx^{\prime }\psi ^{+}(x)\psi ^{+}(x^{\prime })%
\mathcal{V}_{0}(x,x^{\prime })\psi (x)\psi (x^{\prime })  \label{SDFlsdACT}
\end{eqnarray}%
where the integration over the spatial variables is performed over $\Omega
_{C}.$ Here $\mathcal{G}_{0}(x,x^{\prime })$ or its Fourier transform $%
\mathcal{G}_{0}(\mathbf{r},\mathbf{r}^{\prime },i\omega )$ is identified as
the bath Green function appeared in the Dyson equation for the local mass
operator $\mathcal{M}_{int}(\mathbf{r},\mathbf{r}^{\prime },i\omega )$ and
for the local Green function $G_{loc}(\mathbf{r},\mathbf{r}^{\prime
},i\omega )$ of the cluster, and $\mathcal{V}_{0}(\mathbf{r},\mathbf{r}%
^{\prime },i\omega )$ is the "bath interaction" appeared in the Dyson
equation for the local susceptibility $\mathcal{P}(\mathbf{r},\mathbf{r}%
^{\prime },i\omega )$ and local interaction $W_{loc}(\mathbf{r},\mathbf{r}%
^{\prime },i\omega ),$ i.e%
\begin{eqnarray}
\mathcal{G}_{0}^{-1}(\mathbf{r},\mathbf{r}^{\prime },i\omega )
&=&G_{loc}^{-1}(\mathbf{r},\mathbf{r}^{\prime },i\omega )+\mathcal{M}_{int}(%
\mathbf{r},\mathbf{r}^{\prime },i\omega )  \label{SDFlsdG01} \\
\mathcal{V}_{0}^{-1}(\mathbf{r},\mathbf{r}^{\prime },i\omega )
&=&W_{loc}^{-1}(\mathbf{r},\mathbf{r}^{\prime },i\omega )+\mathcal{P}(%
\mathbf{r},\mathbf{r}^{\prime },i\omega )  \label{SDFlsdV01}
\end{eqnarray}%
Note that neither $\mathcal{G}_{0}$ nor $\mathcal{V}_{0}$ can be associated
with non--interacting $G_{0}$ and bare interaction $v_{C},$ respectively.
Note also that both $\mathbf{r}$ and $\mathbf{r}^{\prime }$ indexes in $%
\mathcal{G}_{0}(\mathbf{r},\mathbf{r}^{\prime },i\omega )$ and in $\mathcal{V%
}_{0}(\mathbf{r},\mathbf{r}^{\prime },i\omega )$ vary within the cellular
area $\Omega _{C}.$ The same should be assumed for the local quantities $%
G_{loc}(\mathbf{r},\mathbf{r}^{\prime },i\omega )$ and $W_{loc}(\mathbf{r},%
\mathbf{r}^{\prime },i\omega ).$ Since these functions are translationally
invariant on the original lattice, this property can be used to set up these
functions within $\Omega _{C}.$

An interesting observation can be made on the role of the impurity model
which in the present context appeared as an approximate way to extract the
self--energy of the lattice using input bath Green function and bath
interaction. Alternatively, the impurity problem can be thought itself as
the model which delivers exact mass operator of the spectral density
functional \cite{Chitra1}. If the latter is known, there should exist such
bath Green function and such bath interaction which can be used to reproduce
it. In this respect, the local interaction $W_{loc}$ appeared in our
formulation can be thought as an exact way to define the local Coulomb
repulsion "$U$", i.e. such interaction which delivers exact local
self--energy.

\begin{figure}[tbh]
\includegraphics*[height=2.5in]{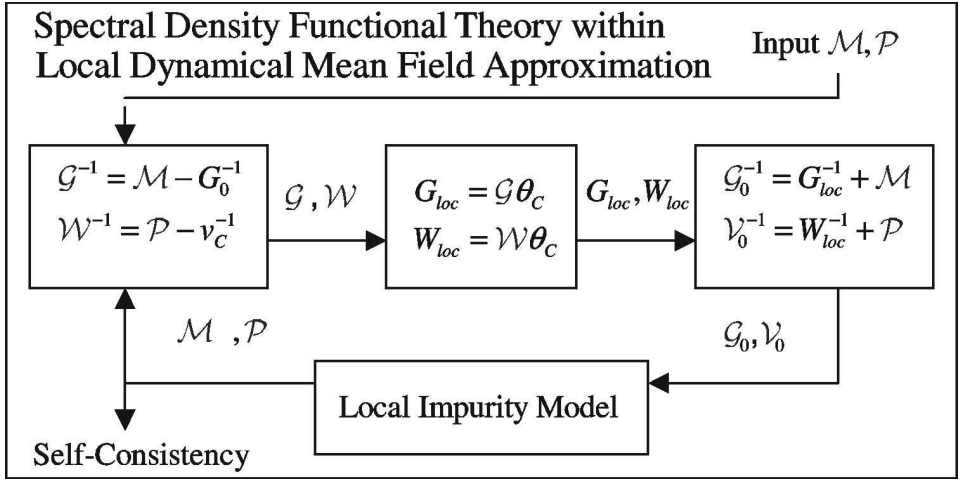}
\caption{Illustration of self-consistent cycle in spectral density
functional theory with local dynamical mean-field approximation: both local
Green function $G_{loc}$ and local Coulomb interaction $W_{loc}$ are
iterated. The auxiliary quantities $\mathcal{G}$ and $\mathcal{W}$ are used
to simplify the construction of the functional.}
\label{FigSDFT}
\end{figure}

To summarize, the effective action for the cluster cell (\ref{SDFlsdACT}),
the Dyson equations (\ref{SDFlsdG01}), (\ref{SDFlsdV01}) connecting local
and bath quantities as well as the original Dyson equations (\ref{SDFactC01}%
), (\ref{SDFactWM1}) constitute a self--consistent set of equations as the
saddle--point conditions extremizing the spectral density functional $\Gamma
_{SDF}(\mathcal{G},\mathcal{W})$. They combine cellular and extended
versions of DMFT and represent our philosophy in the \textit{ab initio}
simulation of a strongly correlated system. Since $\mathcal{M}_{int}$ and $%
\mathcal{P}$ are unknown at the beginning, the solution of these equations
assumes self--consistency. First, assuming some initial $\mathcal{M}_{int},$%
and $\mathcal{P}$ the original Dyson equations (\ref{SDFactC01}), (\ref%
{SDFactWM1}) are used to find Green function $\mathcal{G}$ and screened
interaction $\mathcal{W}.$ Second, the Dyson equations for the local
quantities (\ref{SDFlsdG01}), (\ref{SDFlsdV01}) are used to find $\mathcal{G}%
_{0}$, $\mathcal{V}_{0}.$ Third, quantum impurity model with the cluster
action $S_{loc}$ after (\ref{SDFlsdACT}) is solved by available many--body
technique to give new local $\mathcal{M}_{int}$ and $\mathcal{P}$. The
process is repeated till self--consistency is reached. This is schematically
illustrated in Fig. \ref{FigSDFT}. Note here that while single--site
impurity problem has a well--defined algorithm to extract the lattice
self--energy, this is not generally true for the cluster impurity models 
\cite{Biroli}. The latter provides the self--energy of the cluster, and an
additional prescription such as implemented within cellular DMFT or using
DCA should be given to construct the self--energy of the lattice.

Unfortunately, writing down the precise functional form for $\Phi
_{SDF}[G_{loc},W_{loc}]$ or $\Psi _{SDF}[G_{loc},W_{loc}]$ is still a
problem because the\ evaluation of the entropy requires the evaluation of
the energy as a function of temperature and an additional integration over
it. In general, the free energy $F=E-TS,$ where $E$ is the total energy and $%
S$ is the entropy. Since $\Gamma _{SDF}[\mathcal{G}]=K_{SDF}[\mathcal{G}%
]+\Phi _{SDF}[G_{loc}],$ both energy and entropy terms exist in the kinetic
and interaction functionals. The energy part of $K_{SDF}[\mathcal{G}]=%
\mathrm{Tr(}-\nabla ^{2}+V_{ext})\mathcal{G}\ $and the energy part of $\Phi
_{SDF}[G_{loc},W_{loc}]$ can be written explicitly as $\frac{1}{2}\mathrm{Tr}%
\mathcal{M}_{int}G_{loc}.$ The entropy correction is a more difficult one.
In principle, it can be evaluated by performing calculations of the total
energy $E_{SDF}[\mathcal{G}]=\mathrm{Tr(}-\nabla ^{2}+V_{ext})\mathcal{G}+%
\frac{1}{2}\mathrm{Tr}\mathcal{M}_{int}G_{loc}$ at several temperatures and
then taking the integral \cite{ReviewDMFT}:%
\begin{equation}
S(T)=S(\infty )-\int_{T}^{\infty }dT^{\prime }\frac{1}{T^{\prime }}\frac{%
dE_{SDF}}{dT^{\prime }}  \label{SDFlsdENT}
\end{equation}%
The infinite temperature limit $S(\infty )$ for a well defined model
Hamiltonian can be worked out. This program was implemented for the Hubbard
model \cite{Marcelo} and for\ Ce \cite{CeMcMahan}.

Two well separate problems are now seen. For a given material using the
formulae (\ref{SDFactC01}), (\ref{SDFactWM1}), (\ref{SDFLOC}), (\ref%
{SDFactWLC}), (\ref{SDFlsdG01}), (\ref{SDFlsdV01}) $\mathcal{G},\mathcal{W}%
,G_{loc},W_{loc},\mathcal{G}_{0},\mathcal{V}_{0}$ should be computed using
the methods and algorithms of the electronic structure theory. This
procedure will in part be described in Section III. With given input $%
\mathcal{G}_{0}$ and $\mathcal{V}_{0},$ the solution of the impurity model
constitutes a well separated problem which can be carried out either using
the QMC method or other impurity solver. Some of the techniques are
discussed in Refs. \onlinecite{ReviewDMFT,ReviewLDA+DMFT}. In Section IV,
while applying a simplified version of the theory to plutonium, we will
briefly describe an impurity solver used in that calculation. A full
description of this method will appear elsewhere \cite{Udo}.

The described algorithm is quite general, totally \textit{ab initio} and
allows to determine all quantities, such as the one--electron local Green
functions $G_{loc}$ and the dynamically screened local interactions $W_{loc}$%
. Unfortunately, its full implementation is a very challenging project which
so far has only been carried out at the level of model Hamiltonians\cite%
{Ping}. There are several simplifications which can be made, however. The
screened Coulomb interaction $\mathcal{W}(\mathbf{r},\mathbf{r}^{\prime
},i\omega )$ can be treated on different levels of approximations. In many
cases used in practical calculations with the LDA+DMFT method, this
interaction $\mathcal{W}$ is assumed to be static and parametrized by a set
of some optimally screened on--site parameters, such as Hubbard $U$ and
exchange $J$. These parameters can be fixed by constrained density
functional calculations, extracted from atomic spectra data or adjusted to
fit the experiment. Since the described theory can perform a search in a
constrained space with fixed interaction $\mathcal{W},$ this justifies the
use of $U$ and $J$ as input numbers. A more refined approximation, can use a
method such as GW to generate an energy--dependent $\mathcal{W}$\ \cite%
{Ferdi} which is then treated using extended DMFT \cite{Ping}.
Alternatively\ we can envision that $\mathcal{W}$ is already so short ranged
that we can ignore the EDMFT self consistency condition, and we treat $%
\mathcal{W}$ as $\mathcal{W}_{fix}(x,x^{\prime })$ .This leads to performing
a partial self--consistency with respect to the Green function only. The
procedure is reduced to solving Dyson equations (\ref{SDFactC01}), (\ref%
{SDFlsdG01}) as well as to finding $\mathcal{M}_{int}$ via the solution of
the impurity problem$.$ A full self--consistency can finally be restored by
including a second loop to relax $\mathcal{W}.$

A methodological comment should be made in order to make contact with the
literature of cluster extensions of single site DMFT within model
Hamiltonians. We adopted a less restrictive notion of locality by defining
an effective action of the one particle Green function (and of the
interaction) whose arguments are in nearby unit cells. This maintains the
full translation invariance of the lattice. At the level of the exact
effective action , this is an exact construction, and its extremization will
lead to portions of the exact Greens function which obeys causality. Notice
however that it has been proved recently\cite{Biroli} \ that \textit{%
generating approximations} to the exact functional by restricting the Baym
Kadanoff functional to non local Greens functions leads to violations of
causality. For this reason, we propose to use techniques such as CDMFT which
are manifestly causal for the purpose of realizing approximations to the
local Greens functions.

Our final general comment concerns the optimal choice of local
representation or, precisely, the definition of the local Green function.
This is because the local dynamical mean--field approximation is likely to
be accurate only \ if we know in which portion of the Hilbert space the real
electronic self--energy is well localized. Unfortunately, this is not known 
\textit{a priori}, and in principle, only a full cluster DMFT calculation is
capable to provide us some hints in attempts to answer this question.
However, considerable empirical evidence can be used as a guide for choosing
a basis for DMFT\ calculations, and we discuss these issues in the following
sections.

\subsection{Choice of Local Representation}

We have already pointed out that spectral density functional theory is a
basis set dependent theory since it probes the Green function locally in a
certain region determined by a choice of basis functions in the Hilbert
space. Provided the calculation is exact, the free energy of the system and
the local spectral density in that Hilbert space will be recovered
regardless the choice of it.\ We have developed the theory assuming that the
basis in the Hilbert space is indeed the real space which gives us the
choice (\ref{SDFLOC}) for the local Green function, i.e. the part of the
real Green function restricted by $\theta _{loc}(\mathbf{r},\mathbf{r}%
^{\prime }).$ While this is most natural choice for the purpose of
formulating locality in $\mathbf{r}$ and $\mathbf{r}^{\prime }$ variables,
it is also very useful to discuss a more general choice, connected to some
space of orbitals $\chi _{\xi }(\mathbf{r})$ which can be used to represent
all the relevant quantities in our calculation. As we have in mind to
utilize sophisticated basis sets of modern electronic structure
calculations, we will sometimes waive the orthogonality condition and will
introduce the overlap matrix $O_{\xi \xi ^{\prime }}=\langle \chi _{\xi
}|\chi _{\xi ^{\prime }}\rangle $ especially in cases when we discuss a
practical implementation of the method.

We note that the space $\chi _{\xi }(\mathbf{r})$ can in principle be
interpreted as the reciprocal space plane wave representation $\chi _{\xi }(%
\mathbf{r})=e^{i(\mathbf{k}+\mathbf{G})\mathbf{r}},\xi =\mathbf{k}+\mathbf{G}
$ with $\mathbf{k}$ being the Brillouin zone vector and $\mathbf{G}$ being
the reciprocal lattice vector. Thus the Green function can be probed in the
region of the reciprocal space. It can be interpreted as the real space
representation if $\chi _{\xi }(\mathbf{r})=\delta (\mathbf{\xi }-\mathbf{r}%
) $ where the sums over $\xi $ are interpreted as the integrals over the
volume, and the locality in this basis is precisely exploited in (\ref%
{SDFLOC}). A tremendous transparency of the theory will also arrive if we
interpret the orbital space $\{\chi _{\xi }\}$ as a general non--orthogonal
tight-binding basis set when index $\xi $ combines the angular momentum
index $lm$, and the unit cell index $\mathbf{R},$ i.e. $\chi _{\xi }(\mathbf{%
r})=\chi _{lm}(\mathbf{r}-\mathbf{R})=\chi _{\alpha }(\mathbf{r}-\mathbf{R}%
). $ Note that we can add additional degrees of freedom to the index $\alpha 
$ such, for example, as multiple kappa basis sets of the linear muffin--tin
orbital based methods, Gaussian decay constants in the Gaussian orbital
based methods, and so on. If more than one atom per unit cell is considered,
index $\alpha $ should be supplemented by the atomic basis position within
the unit cell, which is currently omitted for simplicity. For spin
unrestricted calculations $\alpha $ accumulates the spin index $\sigma $ and
the orbital space is extended to account for the eigenvectors of the Pauli
matrix.

Let us now introduce the representation for the exact Green function in the
localized orbital representation%
\begin{equation}
G(\mathbf{r},\mathbf{r}^{\prime },i\omega )=\sum_{\alpha \beta }\sum_{%
\mathbf{k}}\chi _{\alpha }^{\mathbf{k}}(\mathbf{r})G_{\alpha \beta }(\mathbf{%
k},i\omega )\chi _{\beta }^{\mathbf{k}\ast }(\mathbf{r}^{\prime
})=\sum_{\alpha \beta }\sum_{RR^{\prime }}\chi _{\alpha }(\mathbf{r}-\mathbf{%
R})G_{\alpha \beta }(\mathbf{R}-\mathbf{R}^{\prime },i\omega )\chi _{\beta
}^{\ast }(\mathbf{r}^{\prime }-\mathbf{R}^{\prime })  \label{SDFlocREP}
\end{equation}%
Assuming the single--site impurity case, we can separate local and
non--local parts $G_{loc}(\mathbf{r},\mathbf{r}^{\prime },i\omega
)+G_{non-loc}(\mathbf{r},\mathbf{r}^{\prime },i\omega )$ as follows 
\begin{equation}
G_{loc}(\mathbf{r},\mathbf{r}^{\prime },i\omega )=\sum_{\alpha \beta
}G_{loc,\alpha \beta }(i\omega )\sum_{R}\chi _{\alpha }(\mathbf{r}-\mathbf{R}%
)\chi _{\beta }^{\ast }(\mathbf{r}^{\prime }-\mathbf{R})=\sum_{\alpha \beta
}G_{loc,\alpha \beta }(i\omega )\sum_{\mathbf{k}}\chi _{\alpha }^{\mathbf{k}%
}(\mathbf{r})\chi _{\beta }^{\mathbf{k}\ast }(\mathbf{r}^{\prime })
\label{SDFlocGLC}
\end{equation}%
where we denoted the site--diagonal matrix elements $\delta _{RR^{\prime
}}G_{\alpha \beta }(\mathbf{R}-\mathbf{R}^{\prime },i\omega )$ as $%
G_{loc,\alpha \beta }(i\omega )$. Note that this definition is different
from the real--space definition (\ref{SDFLOC}). For example, (\ref{SDFLOC})
contains the information about the density of the system. The formula (\ref%
{SDFlocGLC}) does not describe the density since $\mathbf{R}\neq \mathbf{R}%
^{\prime }$ elements of the matrix $G_{\alpha \beta }(\mathbf{R}-\mathbf{R}%
^{\prime },i\omega )$ are thrown away. The locality of (\ref{SDFlocGLC}) is
controlled exclusively by the decay of the orbitals $\chi _{\alpha }(\mathbf{%
r})$ as a function of $\mathbf{r}$, not by $\theta _{loc}(\mathbf{r},\mathbf{%
r}^{\prime })$

The local part of the Green function, $G_{loc}(\mathbf{r},\mathbf{r}^{\prime
},i\omega ),$ which is just defined with respect to the Hilbert space $%
\{\chi _{\alpha }\}$ can be found by developing the corresponding spectral
density functional theory. Since the basis set is assumed to be fixed, the
matrix elements $G_{loc,\alpha \beta }(i\omega )$ appear only as variables
of the functional. As above, we introduce an auxiliary Green function $%
\mathcal{G}_{\alpha \beta }(\mathbf{k},i\omega )$ to deal with kinetic
energy counterpart. Stationarity yields the matrix equation:%
\begin{equation}
G_{0,\alpha \beta }^{-1}(\mathbf{k},i\omega )=\mathcal{G}_{\alpha \beta
}^{-1}(\mathbf{k},i\omega )+\mathcal{M}_{int,\alpha \beta }(i\omega )
\label{SDFlocDEQ}
\end{equation}%
where the non--interacting Green function (\ref{SDFactG01}) is the matrix of
non--interacting one--electron Hamiltonian%
\begin{equation}
G_{0,\alpha \beta }^{-1}(\mathbf{k},i\omega )=\langle \chi _{\alpha }^{%
\mathbf{k}}|i\omega +\mu +\nabla ^{2}-V_{ext}|\chi _{\beta }^{\mathbf{k}%
}\rangle  \label{SDFlocNON}
\end{equation}%
The self--energy $\mathcal{M}_{int,\alpha \beta }(i\omega )$ is the
derivative $\delta \Phi _{SDF}[G_{loc,\alpha \beta }(i\omega )]/\delta
G_{loc,\alpha \beta }(i\omega )$ and takes automatically the k--independent
form.

While formally exact, this theory would have at least one undesired feature
since, for example, the density of the system can no longer be found from
the definition (\ref{SDFlocGLC}) of $G_{loc}(\mathbf{r},\mathbf{r}^{\prime
},i\omega ).$ As a result the Hartree energy cannot be simply recovered. If
treated exactly $\Phi _{SDF}[G_{loc,\alpha \beta }(i\omega )]$ should
contain the Hartree part. However, we see that the theory delivers
k--independent $\mathcal{M}_{int,\alpha \beta }(i\omega )$ including the
Hartree term. There seems to be a paradox since modern electronic structure
methods calculate the matrix element of the Hartree potential within a given
basis exactly, i.e. $\langle \chi _{\alpha }^{\mathbf{k}}|V_{H}|\chi _{\beta
}^{\mathbf{k}}\rangle .$ The k--dependence is trivial here and is connected
to the known k--dependence of the basis functions used in the calculation.
Therefore, while formulating the spectral density functional theory for
electronic structure calculation, we need to keep in mind that in many
cases, the k--dependence is factorizable and can be brought into the theory
without a problem. This warns us that the choice of the local Green function
has to be done with care so that useful approximations to the functional can
be worked out. It also shows that in many cases the k--dependence is encoded
into the orbitals. It is not that non--trivial k--dependence of the
self--energy operator, which is connected to the fact that $\mathcal{M}%
_{int}\left( \mathbf{r},\mathbf{r}^{\prime },i\omega \right) $ may be
long--range, i.e. decay slowly when $\mathbf{r}$ departs from $\mathbf{r}%
^{\prime }$. It may very well be proportional to $\delta (\mathbf{r}-\mathbf{%
r}^{\prime })$ like the LDA potential and still deliver the k--dependence.

It turn out that the desired k dependence with the choice of the Green
function after (\ref{SDFlocGLC}) can be quickly reinstated if we add the
density of the system as another variable to the functional. This is clear
since the density is a particular case of the local Green function in (\ref%
{SDFLOC} ) taken at $\mathbf{r}=\mathbf{r}^{\prime }$ and summed over $%
i\omega .$ Therefore combination of definition (\ref{SDFlocGLC}) and $\rho $
is another, third possibility of defining $G_{loc}.$ This will allow
treatment of all local Hartree--like potentials without a problem. Moreover,
as we discuss below, this may allow to design better approximations to the
functional since the Hilbert space treatment of locality is more powerful:
it may allow us to treat more long--ranged self--energies than the ones
restricted by $\theta _{loc}(\mathbf{r},\mathbf{r}^{\prime }),$ and the
basis sets can be optimally adjusted to specific self--energies exactly as
the basis sets used in electronic structure calculations are tailored to the
LDA potential.

We have noted earlier that the mass operator $\mathcal{M}_{int}(\mathbf{r},%
\mathbf{r}^{\prime },i\omega )$ is an auxiliary object of the spectral
density functional theory. It has the same meaning as the\ DFT\ Kohn--Sham
potential: it is local self--mass operator that needs to be added to the
non--interacting Green function in order to reproduce the local Green
function of the system, as the DFT potential is added to the
non--interacting Green function to reproduce the density of the system.
Roughly speaking, SDFT provides the \ exact energy and exact one--electron
density of states which is advantageous compared to the DFT which provides
the energy and the density only. However, we obtain \ the full k--dependent
one--particle spectra as the poles of auxiliary Green function $\mathcal{G}(%
\mathbf{r},\mathbf{r}^{\prime },z).$ Can these poles be interpreted as the
exact k--dependent one--electron excitations? This question is similar to
the question of the DFT: can the Kohn--Sham spectra be interpreted as the
physical one--electron excitations? To answer both questions we need to know
something about exact self--energy of the electron. If it is
energy--independent, totally local, i.e. proportional to $\delta (\mathbf{r}-%
\mathbf{r}^{\prime })$ and well--approximated by the DFT\ potential, the
Kohn--Sham spectra represent real one--electron excitations. The exact SDFT
waives most of the restrictions: if the real self--energy is localized
within the area $R_{loc},$ the exact SDFT calculation with the cluster $%
\Omega _{loc}$ including $R_{loc}$ will find the exact k--dependent
spectrum. If we pick $\Omega _{loc}$ larger than $R_{loc},$ the SDFT
equations themselves will choose physical localization area for the
self--energy during our self--consistent calculation. However, these
statements become approximate if we utilize the local dynamical mean field
approximation instead of extremizing the exact functional. Even if the real
self--energy of the electron is sufficiently short--ranged$,$ this
approximation will introduce some error in the calculation, the situation
similar to LDA within DFT. However, the local dynamical mean field theory
does not necessarily have to be formulated in real space. The assumption of
localization for self--energy can be done in some portion of the Hilbert
space. In that portion of the Hilbert space the cluster impurity model needs
to be solved.

The choice of the appropriate Hilbert space, such, e.g., as atomic--like
tight--binding basis set is crucial, to obtain an economical solution of the
impurity model. Let us for simplicity discuss the problem of optimal basis
in some orthogonal tight--binding (Wannier--like) representation for the
electronic self--energy%
\begin{equation}
\Sigma (\mathbf{r},\mathbf{r}^{\prime },i\omega )=\sum_{\alpha \beta }\sum_{%
\mathbf{k}}\chi _{\alpha }^{\mathbf{k}}(\mathbf{r})\Sigma _{\alpha \beta }(%
\mathbf{k},i\omega )\chi _{\beta }^{\mathbf{k}\ast }(\mathbf{r}^{\prime
})=\sum_{\alpha \beta }\sum_{RR^{\prime }}\chi _{\alpha }(\mathbf{r}-\mathbf{%
R})\Sigma _{\alpha \beta }(\mathbf{R}-\mathbf{R}^{\prime },i\omega )\chi
_{\beta }^{\ast }(\mathbf{r}^{\prime }-\mathbf{R}^{\prime })
\label{SDFlocSIG}
\end{equation}%
We can separate our orbital space $\{\chi _{\alpha }\}$ onto the subsets
describing light $\{\chi _{A}\}$ and heavy $\{\chi _{a}\}$ electrons.
Assuming either off--diagonal terms between them are small or we work with
exact Wannier functions, the self--energy $\Sigma (\mathbf{r},\mathbf{r}%
^{\prime },i\omega )$ can be separated onto contributions from the light, $%
\Sigma _{L}(\mathbf{r},\mathbf{r}^{\prime },i\omega ),$ and from the heavy, $%
\Sigma _{H}(\mathbf{r},\mathbf{r}^{\prime },i\omega ),$ electrons. $\Sigma
_{\alpha \beta }(\mathbf{k},i\omega )$ is expected to be k--dependent but
largely $\omega $ independent for the light block, i.e $\Sigma _{L}(\mathbf{r%
},\mathbf{r}^{\prime },i\omega )=\sum {}_{AB}\sum_{\mathbf{k}}\chi _{A}^{%
\mathbf{k}}(\mathbf{r})\Sigma _{AB}(\mathbf{k})\chi _{B}^{\mathbf{k}\ast }(%
\mathbf{r}^{\prime })$. The k--dependency here should be well--described by
LDA--like approximations, therefore we expect $\Sigma _{L}(\mathbf{r},%
\mathbf{r}^{\prime },i\omega )\sim V_{eff}(\mathbf{r})\delta (\mathbf{r}-%
\mathbf{r}^{\prime }).$ A different situation is expected for the heavy
block where we would rely on the result

\begin{equation}
\Sigma _{H}(\mathbf{r,r}^{\prime },i\omega )\sim V_{eff}(\mathbf{r})\delta (%
\mathbf{r}-\mathbf{r}^{\prime })+\sum_{ab}\chi _{a}(\mathbf{r})\Sigma
_{ab}^{\prime }(i\omega )\chi _{b}^{\ast }(\mathbf{r}^{\prime })
\label{SDFlocSTB}
\end{equation}%
The first term here gives the k-dependence coming from an LDA--like
potential. It describes the dispersion in the heavy band. The second term is
the energy dependent correction where site--diagonal approximation $\mathbf{R%
}=\mathbf{R}^{\prime }$ is imposed. What is the best choice of the basis to
use in connection with $\Sigma _{ab}^{\prime }(i\omega )$ in (\ref{SDFlocSTB}%
)? Here the decay of the orbitals $\chi _{\alpha }(\mathbf{r})$ as a
function of $\mathbf{r}$ is now entirely in charge of the self--energy
range. In light of the spectral density functional theory, the answer is the
following: the local dynamical mean field approximation would work best for
such basis $\chi _{\alpha }(\mathbf{r})$ whose range approximately
corresponds to a self--energy localization $R_{loc}$ of the real electron.
Even though $R_{loc}$ is not known \textit{a priori}, something can be
learned about its value based on a substantial empirical evidence. It is,
for example, known that LDA\ energy bands when comparing to experiments at
first place miss the energy dependent $\Sigma _{ab}^{\prime }(i\omega )$
like corrections. This is the case for bandwidths in transition metals (and
also in simple metals), the energy gaps of semiconductors, etc. It is also
known that many--body based theories work best for massively downfolded
model Hamiltonians where only active low--energy degrees of freedom at the
region around the Fermi level $E_{F}$ remain. The many--body Hamiltonian

\begin{equation}
\hat{H}=\sum_{\alpha \beta }\sum_{RR^{\prime }}h_{\alpha R\beta R^{\prime
}}^{(0)}[c_{\alpha R}^{+}c_{\beta R^{\prime }}+h.c.]+\sum_{\alpha \beta
\gamma \delta }\sum_{RR^{\prime }R"R^{\prime \prime \prime }}V_{\alpha \beta
\gamma \delta }^{RR^{\prime }R^{\prime \prime }R^{\prime \prime \prime
}}c_{\alpha R}^{+}c_{\beta R^{\prime }}^{+}c_{\delta R^{\prime \prime \prime
}}c_{\gamma R^{\prime \prime }}  \label{SDFlatHAM}
\end{equation}%
with $V_{\alpha \beta \gamma \delta }^{RR^{\prime }R"R^{\prime \prime \prime
}}=\int d\mathbf{r}d\mathbf{r}^{\prime }\chi _{\alpha R}^{\ast }(\mathbf{r}%
)\chi _{\beta R^{\prime }}^{\ast }(\mathbf{r}^{\prime })v_{C}(\mathbf{r}-%
\mathbf{r}^{\prime })\chi _{\gamma R^{\prime \prime }}(\mathbf{r})\chi
_{\delta R^{\prime \prime \prime }}(\mathbf{r}^{\prime })$ assumes the
one--electron Hamiltonian $h_{\alpha R\beta R}^{(0)}$ is obtained as a fit
to the bands near $E_{F}.$ This can always be done by long ranged Wannier
functions. It is also clear that the correlation effects are important at
first place for the partially occupied bands since only these bring various
configurational interactions in the many--body electronic wave functions.
For example, the well--known one--band Hamiltonian for CuO$_{2}$ plane of
high--T$_{c}$ materials considers an antibonding combination of Cu$%
_{x^{2}-y^{2}}$ and O$_{x,y}$ orbitals which crosses $E_{F}$. Also, the
calculations based on the LDA+DMFT method usually obtain reliable results
when treating only the bands crossing the Fermi level as the correlated
one--electron states. This is, for example, the case of Pu or our \cite%
{OpticsDMFT} and previous \cite{LaTiO3} calculation for LaTiO$_{3}$ where $%
t_{2g}$ three band Hamiltonian is considered. All this implies that the
range for $\Sigma _{ab}^{\prime }(i\omega )$ term in (\ref{SDFlocSTB})
should correspond to the properly constructed Wannier orbitals, \ which is
fairly long--ranged. What happen if we instead utilize mostly localized
representation which, for example, can be achieved by tight--binding fits to
the energy bands at higher energy scale. For the case of CuO$_{2}$ this
would correspond to a three band Hamiltonian with Cu$_{x^{2}-y^{2}}$ and O$%
_{x,y}$ orbitals treated separately. For LaTiO$_{3}$ system this is a
Hamiltonian derived from Ti$_{t2g}$ and O$_{p}$ orbitals. The answer here
can be given as a practical matter of most economic way to solve the
impurity problem: provided Cu and O levels are well separated, provided both
approaches use properly downfolded for each case Coulomb interaction matrix
elements $V_{\alpha \beta \gamma \delta }^{RR^{\prime }R"R^{\prime \prime
\prime }},$ and provided correlations are treated on all orbitals, the final
answer should be similar regardless the choice of the basis. A faster
algorithm will be obtained by treating the one--band Hamiltonian with
antibonding Cu$_{x^{2}-y^{2}}$-- O$_{x,y}$ orbital. If indeed the
self--energy is localized on the scale of the distance between Cu and O, it
is clear where the inefficiency of the three--band model appears: the second
term in (\ref{SDFlocSTB}) needs to be extended within the cluster involving
both Cu and nearest O sites and should involve both Cu and O centered
orbitals simply to reach the cluster boundary. In the one-band case this is
encoded into the decay of the properly constructed Wannier state.

The previous discussion is merely a conjecture. It does not imply that the
localization range for the real self--energy of correlated electron at given
frequency $\omega $ is directly proportional to the size of Wannier states
located in the vicinity of $\omega +\mu .$ It may very well be that in many
cases this range is restricted by a single atom only (atomic sphere of Cu in
the example above). Clearly more experience can be gained by studying a
correlation between the decay of the Coulomb matrix element $V^{RRR^{\prime
}R^{\prime }}$ as a function of $\mathbf{R}-\mathbf{R}^{\prime }$ and the
obtained matrix $\Sigma (\mathbf{R}-\mathbf{R}^{\prime },\omega )$ using a
suitable cluster DMFT technique.\ These works are currently being performed
and will be reported elsewhere\cite{Indranil}. The given discussion however
warns that in general the best choice of the basis for single--site
dynamical mean field treatment may not be the case of mostly localized
representation. In this respect the area restricted by $\theta _{loc}(%
\mathbf{r},\mathbf{r}^{\prime })$ which is used to formulate SDFT in the
real space may need to be extended up to a cluster. However, alternative
formulation with the choice of local Green function after (\ref{SDFlocGLC})
may be more economical since a single--site approximation may still deliver
good results. As we have argued, such spectral density functional theory
will also need the density of the system to complete the definition of local
Green function. The local dynamical mean field approximation can be applied
to the interaction functional $\Phi _{SDF}$ which is viewed as $\Phi
_{SDF}[\rho ,G_{loc}].$ This idea is used by the LDA+DMFT method described
below.

\subsection{LDA+DMFT\ Method}

Various methods such as LDA+U \cite{ReviewLDA+U}, LDA+DMFT\cite%
{ReviewLDA+DMFT} and local GW\cite{Zein,ReviewTsvelik} which appeared
recently for realistic calculations of properties of strongly correlated
materials can be naturally understood within spectral density functional
theory. Let us, for example, explore the idea of expressing the energy as
the density functional. Local density approximation prompts us that a large
portion of the exchange--correlation part $\Phi _{xc}[\rho ]$ can be found
easily. Indeed, the charge density is known to be accurately obtained by the
LDA. Why not think of LDA as the most primitive impurity solver, which
generates manifestly local self--energy with localization radius collapsed
to a single $\mathbf{r}$ point? It is tempting to represent $\Phi
_{SDF}[G_{loc}]=E_{H}[\rho ]+E_{xc}^{LDA}[\rho ]+\tilde{\Phi}[G_{loc}]-\Phi
_{DC}[G_{loc}],$ where the new functional $\tilde{\Phi}_{SDF}[G_{loc}]$
needs in fact to take care of those electrons which are strongly correlated
and heavy, thus badly described by LDA. Conceptually, that means that the
solution of the cluster impurity model for the light electrons is
approximated by LDA and does not need a frequency resolution for their
self--energies.

Unfortunately, the LDA has no diagrammatic representation, and it is
difficult to separate the contributions from the light and heavy electrons.
The $E_{xc}^{LDA}[\rho ]$ is a non--linear functional and it already
includes the contribution to the energy from all orbitals in some average
form. Therefore we need to take care of a non--trivial double counting,
encoded in the functional $\Phi _{DC}[G_{loc}]$. The precise form of the
double counting is related to the approximation imposed for $\tilde{\Phi}%
[G_{loc}]$. We postpone this discussion until establishing the connection to
the LDA+U method in the following subsection.

The LDA+DMFT\ approximation considers both the density and the local Green
function $G_{loc,\alpha \beta }(i\omega )$ defined in (\ref{SDFlocGLC}) as
the parameters of the spectral density functional \cite{SKcondmat}. A
further approximation is made to accelerate the solution of a single--site
impurity model: the functional dependence comes from the subblock of the
correlated electrons only. If localized orbital representation $\{\chi
_{\alpha }\}$ is utilized, a subspace of the heavy electrons $\{\chi _{a}\}$
can be identified. Thus, the approximation can be written as $\tilde{\Phi}%
_{SDF}[G_{loc,ab}(i\omega )]$ , where $G_{loc,ab}(i\omega )$ is the heavy
block of the local Green function$.$ The double counting correction depends
only on the average density of the heavy electrons. Its precise form will be
discussed below, but for now we assume that $\Phi _{DC}[G_{loc}]=\Phi _{DC}[%
\bar{n}_{c}]$ with $\bar{n}_{c}=T\sum_{i\omega }\sum_{a}G_{loc,aa}(i\omega
)e^{i\omega 0^{+}}$, where index $a$ runs within a correlated $l_{c}$ shell
only. We can write the LDA+DFMT\ approximation for the interaction energy as
follows:%
\begin{equation}
\Phi _{LDA+DMFT}[\rho ,G_{loc}]=E_{H}[\rho ]+E_{xc}^{LDA}[\rho ]+\tilde{\Phi}%
[G_{loc,ab}(i\omega )]-\Phi _{DC}[\bar{n}_{c}]  \label{SDFldaLDA}
\end{equation}%
The kinetic energy part is treated as usual with introducing the auxiliary
Green function $\mathcal{G}(\mathbf{r},\mathbf{r}^{\prime },i\omega ).$

The full functional $\Gamma _{LDA+DMFT}[\mathcal{G}]$ is considered as a
functional of the matrix $\mathcal{G}_{\alpha \beta }(\mathbf{R}-\mathbf{R}%
^{\prime },i\omega )$ or its Fourier transformed analog $\mathcal{G}_{\alpha
\beta }(\mathbf{k},i\omega ).$ The stationarity is examined with respect to $%
\mathcal{G}_{\alpha \beta }(\mathbf{k},i\omega )$ and produces the
saddle--point equation similar to (\ref{SDFactC01}). It has the following
matrix form%
\begin{equation}
G_{0,\alpha \beta }^{-1}(\mathbf{k},i\omega )=\mathcal{G}_{\alpha \beta
}^{-1}(\mathbf{k},i\omega )+\mathcal{M}_{int,\alpha \beta }(\mathbf{k}%
,i\omega )  \label{SDFldaDYS}
\end{equation}%
where the non--interacting Green function (\ref{SDFactG01}) is the matrix of
non--interacting one--electron Hamiltonian%
\begin{equation}
G_{0,\alpha \beta }^{-1}(\mathbf{k},i\omega )=\langle \chi _{\alpha }^{%
\mathbf{k}}|i\omega +\mu +\nabla ^{2}-V_{ext}|\chi _{\beta }^{\mathbf{k}%
}\rangle  \label{SDFldaHAM}
\end{equation}%
The self--energy $\mathcal{M}_{int,\alpha \beta }(\mathbf{k},i\omega )$ is
the variational derivative of $\Phi _{LDA+DMFT}[\rho ,G_{loc}].$ Its precise
form depends on the basis set used in the LDA+DMFT calculation.

In general, it can be split onto several contributions including Hartree,
LDA exchange--correlation, DMFT and the double--counting correction. In
orthogonal tight--binding, both DMFT, $\mathcal{\tilde{M}}_{ab}(i\omega ),$
and double counting, $\delta _{ab}V_{aa}^{DC},$ matrices do not depend on $%
\mathbf{k}$. These matrices are non--zero within the heavy block only. The
Dyson equation (\ref{SDFldaDYS}) can be rewritten by separating from $%
\mathcal{M}_{int,\alpha \beta }(\mathbf{k},i\omega )$ the total LDA
potential $V_{LDA}(\mathbf{r})=V_{ext}(\mathbf{r})+V_{H}(\mathbf{r}%
)+V_{xc}^{LDA}(\mathbf{r})$:%
\begin{equation}
\mathcal{G}_{\alpha \beta }^{-1}(\mathbf{k},i\omega )=\langle \chi _{\alpha
}^{\mathbf{k}}|i\omega +\mu +\nabla ^{2}-V_{LDA}|\chi _{\beta }^{\mathbf{k}%
}\rangle -\delta _{\alpha a}\delta _{\beta a}V_{aa}^{DC}+\delta _{\alpha
a}\delta _{\beta b}\mathcal{\tilde{M}}_{ab}(i\omega )  \label{SDFldaGAB}
\end{equation}

The Green function $\mathcal{G}_{\alpha \beta }(\mathbf{k},i\omega )$
obtained from (\ref{SDFactG01}) is used to find $G_{loc,\alpha \beta
}(i\omega )=\sum_{\mathbf{k}}\mathcal{G}_{\alpha \beta }(\mathbf{k},i\omega
) $ which is then used in another Dyson equation to compute the bath Green
function:%
\begin{equation}
\mathcal{G}_{0,ab}^{-1}(i\omega )=G_{loc,ab}^{-1}(i\omega )+\mathcal{\tilde{M%
}}_{ab}(i\omega )  \label{SDFldaG00}
\end{equation}%
In Section III we will also describe an accurate procedure to solve the real
space form (\ref{SDFlsdG01}) of the Dyson equation using the LMTO\ basis
set. The LDA+DMFT bath Green function $\mathcal{G}_{0,ab}(i\omega )$ is the
only essential input to the auxiliary impurity model. Thus, the procedure of
self--consistency within LDA+DMFT\ is reduced to the following steps. First,
some self--energy matrix of the heavy orbitals $\mathcal{\tilde{M}}%
_{ab}(i\omega )$ is guessed. Then, the Dyson equation (\ref{SDFldaDYS}) is
solved in the entire Hilbert space and delivers the Green function $\mathcal{%
G}_{\alpha \beta }(\mathbf{k},i\omega ).$ After that, the local Green
function of the correlated electrons is constructed, which is then used in
the equation (\ref{SDFldaG00}) to deliver the bath Green function $\mathcal{G%
}_{0,ab}(i\omega )$. This matrix is the input to the impurity model.
Solution of this model delivers the new self--energy $\mathcal{\tilde{M}}%
_{ab}(i\omega )$ and the process is iterated towards self--consistency.

\begin{figure}[tbh]
\includegraphics*[height=2.5in]{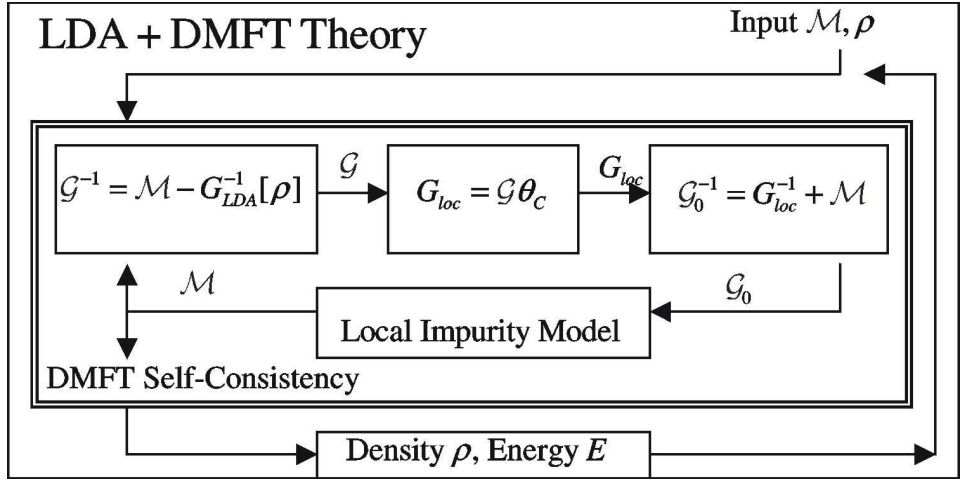}
\caption{Illustration of self-consistent cycle in spectral density
functional theory with LDA+DMFT approximation: double iterational cycle
consists of the innner DMFT loop and outer (density plus total energy) loop.}
\label{FigLDA+DMFT}
\end{figure}

Notice that once the DMFT self--consistency is reached, the process can
either be stopped or continued since the Green function $\mathcal{G}_{\alpha
\beta }(\mathbf{k},i\omega )$ delivers new charge density of the system
which modifies the Hartree and LDA exchange--correlation potentials in the
expression (\ref{SDFldaGAB}). In this respect, the LDA+DMFT method assumes a
double iterational loop, the internal one over the self--energy and the
external one over the density. This is precisely dictated by the spectral
density functional stationarity condition. We illustrate such loop on Fig. %
\ref{FigLDA+DMFT}. Note that in order to access accurate total energies and
remove ambiguity that the LDA Green function (and not any other one) is used
as an input to the DMFT calculation, this density self--consistency loop
needs to be carried out. Our application to the volume expansion in Pu
described later in this paper involves solution of the SDFT equations
allowing the full relaxation of the charge density.

Iterations over the density are not complicated to incorporate in the
programs for electronic structure calculations. The spherical part of the
density at a given site can be written approximately using the atomic sphere
approximation as an integral over the partial density of states $N_{l}(E)$
till the Fermi level $E_{F}$%
\begin{equation}
\rho (r)=\sum_{l}\int_{-\infty }^{E_{F}}N_{l}(E)\varphi _{l}^{2}(r,E)dE
\label{SDFldaRHO}
\end{equation}%
where $\varphi _{l}(r,E)$ are the solutions of the radial Schr\"{o}dinger
equation. Usually these are constructed using spherical part of the LDA
potential but in the present context the non--local self--energy operator
needs to be utilized\cite{ZeinLGW}:

\begin{equation}
(-\nabla _{rl}^{2}-E)\varphi _{l}(r,E)+\int \mathcal{M}_{eff}(r,r^{\prime
},E)\varphi _{l}(r^{\prime },E)r^{2}dr=0  \label{SDFldaPHI}
\end{equation}%
Expression (\ref{SDFldaRHO}) can be simplified further if we assume a Taylor
expansion of $\varphi _{l}(r,E)$=$\varphi _{l}(r,E_{\nu l})+(E-E_{\nu l})%
\dot{\varphi}_{l}(r,E_{\nu l})$ around some linearization energies $E_{\nu
l} $ taken at the centers of gravities of the occupied energy bands, i.e. $%
E_{\nu l}=\int_{-\infty }^{E_{F}}EN_{l}(E)dE/\int_{-\infty
}^{E_{F}}N_{l}(E)dE$. During the iterations, DMFT modifies the density of
states, $\Delta N_{l}(E),$ which leads to the change $\Delta \rho (r)$ of
the density. The latter has a feedback onto the change $\Delta \varphi
_{l}(r,E)$ [or changes $\Delta \varphi _{l}(r,E_{\nu l}),\Delta \dot{\varphi}%
_{l}(r,E_{\nu l})$]. If we assume that these changes are small comparing to
the original LDA values, we can work out a linear response relationship for
the density%
\begin{equation}
\Delta \rho (r)=\sum_{l}\int_{-\infty }^{E_{F}}\Delta N_{l}(E)\varphi
_{l}^{2}(r,E)dE+\sum_{l}\int_{-\infty }^{E_{F}}N_{l}(E)\varphi
_{l}(r,E)\Delta \varphi _{l}(r,E)dE+\sum_{l}N_{l}(E_{F})\varphi
_{l}^{2}(r,E_{F})\Delta E_{F}  \label{SDFldaDRO}
\end{equation}%
and for the LDA potential%
\begin{equation}
\Delta V_{LDA}(r)=e^{2}\int \frac{\Delta \rho (r^{\prime })}{|\mathbf{r}-%
\mathbf{r}^{\prime }|}d\mathbf{r}^{\prime }+\frac{dV_{xc}^{LDA}}{d\rho }%
\Delta \rho (r)  \label{SDFldaPOT}
\end{equation}%
Thus, to first order, these are the quantities which are iterated in the
external density loop of the spectral density functional as shown on Fig. %
\ref{FigLDA+DMFT}.

The main physical point of the LDA+DMFT methodology\ is identification of a
subset of the correlated orbitals $\{\chi _{a}\}$ which is separated from
the full Hilbert space $\{\chi _{\alpha }\}.$ In the case like Pu, this is
the subset of f--electron orbitals. In other situations, this subset can be
isolated based on physical grounds. If $\{\chi _{a}\}$ is appropriately
constructed Wannier representation, this subset may describe the bands
crossing the Fermi level. We expect the dynamical self--energy corrections
to appear at first place only within the subset $\{\chi _{a}\}$. However,
changes in the electronic densities of states, $\Delta N_{l}(E),$ will
appear for all light and heavy electrons.

We did not discuss so far the relaxation of the screened Coulomb interaction 
$\mathcal{W}(\mathbf{r},\mathbf{r}^{\prime },i\omega )$, which, in
principle, needs to be done during the self--consistency in parallel to $%
\mathcal{G}(\mathbf{r},\mathbf{r}^{\prime },i\omega )$. We stress that the
short--range behavior appears only for the local effective susceptibility $%
\mathcal{P}(\mathbf{r},\mathbf{r}^{\prime },i\omega )$ in (\ref{SDFactDEW})
and means its fast decay when $\mathbf{r}$ departs from $\mathbf{r}^{\prime
}.$ Contrary, the function $\mathcal{W}(\mathbf{r},\mathbf{r}^{\prime
},i\omega )$ can be as long range as the bare Coulomb interaction if
necessary. This is dictated by Eq. (\ref{SDFactDEW}) and is similar to the
relationship (\ref{SDFactC01}) between $\mathcal{G}$ and $\mathcal{M}_{int}.$

The locality \ assumption for $\mathcal{P}$ should simplify the
self--consistency over $\mathcal{W}$. This should be faster than the one
employed in the full GW method which formally tries to compute full $\Pi (%
\mathbf{r},\mathbf{r}^{\prime },i\omega ).$ In the language of local orbital
representation $\{\chi _{\alpha }\},$ this means computation of all matrix
elements for $\Pi _{\alpha \beta }(\mathbf{R}-\mathbf{R}^{\prime },i\omega )$
or its Fourier transform $\Pi _{\alpha \beta }(\mathbf{k},i\omega )$ as
compared to the site--diagonal ($\delta _{RR^{\prime }})$ or small cluster
cases of SDFT. This will be discussed below in connection to the recently
proposed \cite{Zein,ReviewTsvelik} local version of the GW method.

So far we did not mention the problem of the optimal choice of the double
counting corrections. This is discussed below in connection to the LDA+U\
method.

\subsection{Double Counting and LDA+U Method}

Historically, the LDA+U\ method has been introduced \cite{AnisimovLDA+U} as
an extension of the local spin density approximation (LSDA) to treat the
ordered phases of Mott insulating solids. In this respect it is a natural
extension of LSDA. However, this method was first to recognize that a better
energy functional can be constructed if not only the density but the density
matrix of correlated orbitals is brought into the density functional. We
have discussed the correlated subset $\{\chi _{a}\}$ and local Green
functions $G_{loc,\alpha \beta }(i\omega )$ in connection to the LDA+DMFT
method. The density matrix $n_{ab}$ is related to the correlated subblock of
the local Green function%
\begin{equation}
n_{ab}=T\sum_{i\omega }e^{i\omega 0^{+}}G_{loc,ab}(i\omega )
\label{SDFlduNAB}
\end{equation}%
Therefore, the LDA+U method can be viewed as an approximation
(Hartree--Fock\ approximation) to the spectral density functional within
LDA+DMFT.

The correct interaction energy among the correlated electrons can be written
down explicitly using the Hartree--Fock approximation. In our language the
LDA+DMFT\ interaction energy functional (\ref{SDFldaLDA}) is rewritten in
the form

\begin{equation}
\Phi _{LDA+U}[\rho ,n_{ab}]=E_{H}[\rho ]+\Phi _{xc}^{LDA}[\rho ]+\tilde{\Phi}%
_{U}[n_{ab}]-\Phi _{DC}[\bar{n}_{c}]  \label{SDFlduFUN}
\end{equation}%
where the functional form $\tilde{\Phi}_{U}[n_{ab}]$ is known explicitly: 
\begin{equation}
\tilde{\Phi}_{U}[n_{ab}]=\frac{1}{2}%
\sum_{abcd}(U_{acbd}-U_{acdb})n_{ab}n_{cd}  \label{SDFlduFHF}
\end{equation}%
Here, indexes $a,b,c,d$ involve fixed angular momentum $l$ of the heavy
orbitals and run over magnetic $m$ and spin $\sigma $ quantum numbers. The
on--site Coulomb interaction matrix $U_{abcd}$ is the on--site Coulomb
interaction matrix element $V_{\alpha =a\beta =b\gamma =c\delta =d}^{RRRR}$
appeared in (\ref{SDFlatHAM}) which is again taken for the subblock of the
heavy orbitals. Note that sometimes $U_{abcd}$ is defined as $V_{\alpha
=a\beta =c\gamma =b\delta =d}^{RRRR}.$

The double counting term $\Phi _{DC}[n_{ab}]$ needs to be introduced since
both the L(S)DA and U terms account for the same interaction energy between
the correlated orbitals. This includes in first place the Hartree part.
However, the precise form of the double counting is unclear due to
non--linear nature of the LDA exchange--correlation energy. In practice, it
was proposed\cite{ReviewLDA+U} that the form for $\Phi _{DC}$ is

\begin{equation}
\Phi _{DC}[\bar{n}_{c}]=\frac{1}{2}\bar{U}\bar{n}_{c}(\bar{n}_{c}-1)-\frac{1%
}{2}\bar{J}[\bar{n}_{c}^{\uparrow }(\bar{n}_{c}^{\uparrow }-1)+\bar{n}%
_{c}^{\downarrow }(\bar{n}_{c}^{\downarrow }-1)].  \label{SDFlduFDC}
\end{equation}%
where $\bar{U}=\frac{1}{(2l+1)^{2}}\sum_{ab}U_{abab}$ , $\bar{J}=\bar{U}-%
\frac{1}{2l(2l+1)}\sum_{ab}(U_{abab}-U_{abba})$and where $\bar{n}%
_{c}^{\sigma }=\sum_{a\in l_{c}}n_{aa}\delta _{\sigma _{a}\sigma },$ $\bar{n}%
_{c}=\bar{n}_{c}^{\uparrow }+\bar{n}_{c}^{\downarrow }.$ Some other forms of
the double countings have also been discussed in Ref. \onlinecite{Mazin}.

The minimization of the functional $\Gamma _{LDA+U}[\rho ,n_{ab}]$ is now
performed. The self--energy correction in (\ref{SDFldaGAB}) appears as the
orbital dependent correction $\mathcal{\tilde{M}}_{ab}-V_{ab}^{DC}$:%
\begin{eqnarray}
\mathcal{\tilde{M}}_{ab} &=&\frac{\delta \tilde{\Phi}_{U}}{\delta n_{ab}}%
=\sum_{cd}(U_{acbd}-U_{acdb})n_{cd}  \label{SDFlduSIG} \\
V_{ab}^{DC} &=&\frac{\delta \Phi _{DC}}{\delta n_{ab}}=\delta _{ab}[\bar{U}(%
\bar{n}_{c}-\frac{1}{2})-\bar{J}(\bar{n}_{c}^{\sigma }-\frac{1}{2})]
\label{SDFlduSDC}
\end{eqnarray}%
While the correction is static, it is best viewed as the Hartree--Fock
approximation to the self--energy $\mathcal{M}_{ab}(i\omega )$ within the
LDA+DMFT method. Note that such interpretation allows us to utilize double
counting forms within LDA+DMFT as $\mathcal{\tilde{M}}(\mathbf{r},\mathbf{r}%
^{\prime },i\infty )$ or $\mathcal{\tilde{M}}(\mathbf{r},\mathbf{r}^{\prime
},i0)$. Note also that the solution of the impurity problem collapses in the
LDA+U method since the self--energy is known analytically by formula (\ref%
{SDFlduSIG}).

From a practical point of view, despite the great success of the LDA+U
theory in predicting materials properties of correlated solids \cite%
{ReviewLDA+U} there are obvious problems with this approach when applied to
metals or to systems where the orbital symmetries are not broken. They stem
from the well--known deficiencies of the Hartree--Fock approximation. The
most noticeable is that it only describes spectra of magnetically ordered
systems which have Hubbard bands. We have however argued that a correct
treatment of the electronic structure of strongly correlated systems has to
treat both Hubbard bands and quasiparticle bands on the same footing.
Another problem occurs in the paramagnetic phase of Mott insulators: in the
absence of any broken symmetry the LDA+U method reduces to the LDA, and the
gap collapses. In systems like NiO where the gap is of the order of eV, but
the Neel temperature is a few hundred Kelvin, it is unphysical to assume
that the gap and the magnetic ordering are related. For this reason the
LDA+U predicts magnetic order in cases that it is not observed, as, e.g., in
the case of Pu \cite{PuPRL}.

\subsection{Local GW\ Approximation}

We now discuss the relaxation of the screened Coulomb interaction $\mathcal{W%
}(\mathbf{r},\mathbf{r}^{\prime },i\omega )$ which appeared in the spectral
density functional formulation of the problem. Both LDA+DMFT and LDA+U
methods parametrize the interaction $\mathcal{W}$ with optimally screened
set of parameters, such, e.g, as the matrix $U_{abcd}$ appeared in (\ref%
{SDFlduFHF}). This matrix is supposed to be given by an external calculation
such, e.g., as the constrained LDA method \cite{ConstrainedDFT}. To
determine this interaction self consistently an additional self--consistency
loop described by the equations (\ref{SDFactDEW}) and (\ref{SDFlsdV01}) has
to be switched on together with calculation of the local susceptibility $%
\mathcal{P}(\mathbf{r},\mathbf{r}^{\prime },i\omega )$ by the impurity
solver. This brings a truly self--consistent \textit{ab initio} method
without input parameters and the double counting problems.

A simplified version of this method has been recently proposed \cite%
{ReviewTsvelik,Zein} which is known as a local version of the GW method
(LGW). Within the spectral density functional theory, this approximation
appears as approximation to the functional $\Psi _{SDF}[G_{loc},W_{loc}]$
taken in the form%
\begin{equation}
\Psi _{LGW}[G_{loc},W_{loc}]=-\frac{1}{2}\mathrm{Tr}G_{loc}W_{loc}G_{loc}
\label{SDFlgwLOC}
\end{equation}%
As a result, the susceptibility $\mathcal{P}(\mathbf{r},\mathbf{r}^{\prime
},i\omega )$ is approximated by the product of two local Green functions,
i.e.$\mathcal{P}=-2\delta \Psi _{LGW}/\delta W_{loc}=G_{loc}G_{loc},$ and
the exchange--correlation part of our mass operator is approximated by the
local GW diagram, i.e. $\mathcal{M}_{xc}=\delta \Psi _{LGW}/\delta
G_{loc}=-G_{loc}W_{loc}$.

Thus, the impurity model is solved and the procedure can be made
self--consistent: For a given $\mathcal{M}_{int}$ and $\mathcal{P}$, the
Dyson equations (\ref{SDFactC01}), (\ref{SDFactWM1}) for $\mathcal{G}$ and $%
\mathcal{W}$ are solved. Then, the local quantities $G_{loc,}$ $W_{loc}$ are
generated and used to find new $\mathcal{M}_{int}$ and $\mathcal{P}$ thus
avoiding the computation of the bath Green function $\mathcal{G}_{0}$ after (%
\ref{SDFlsdG01}), and the interaction $\mathcal{V}$, after (\ref{SDFlsdV01}).

Note that since the local GW approximation (\ref{SDFlgwLOC}) is relatively
cheap from computational point of view, its implementation on a cluster and
for all orbitals should not be a problem. The results of the single--site
approximation for the local quantities have been developed independently and
reported in the literature.\cite{Zein}. The cluster extension is currently
being performed and the results will be reported elsewhere \cite{ZeinLGW}.

Note finally that the local GW approximation is not the only one which can
be implemented as the simplified impurity solver. For example, another
popular approximation known as the fluctuational exchange approximation
(FLEX)\ can be worked out along the same lines. Note also that the
combination of the DMFT and full GW\ diagram has been recently proposed \cite%
{ReviewTsvelik,Georges} and a simplified implementation for Ni \cite{Georges}%
, and for a model Hamiltonian \cite{Ping} have been carried out. This
procedure incorporates full k--dependence of the self--energy known
diagrammatically within GW together with the additional local DMFT\ diagrams.

\section{Calculation of Local Green Function}

The solution of the Dyson equations described in the previous section for a
given strongly correlated material requires the calculation of the local
Green function during the iterations towards self--consistency. This is very
similar to the procedure in the density functional theory, when the charge
density is computed. A big advantage of DFT is the use of Kohn--Sham
orbitals which reduces the equation (\ref{SDFactKS1}) for the Kohn--Sham
Green function to a set of one--particle Schr\"{o}dinger's like equations
for the wave functions. As a result the kinetic energy contribution is
calculated directly and the evaluation of the total energy of a solid is not
a problem. Here, a similar algorithm will be described for the
energy--dependent Dyson equation, the solution in terms of the
linear--muffin--tin orbital basis set will be discussed, and the formula for
evaluating the total energy will be given.

\subsection{Energy Resolved One-Particle Representation}

We introduced the auxiliary Green function $\mathcal{G}(\mathbf{r},\mathbf{r}%
^{\prime },i\omega )$ to deal with the kinetic part of the action in SDFT.
It satisfies to the Dyson equation (\ref{SDFactG01}). Let us now introduce
the representation of generalized energy--dependent one-particle states

\begin{eqnarray}
\mathcal{G}(\mathbf{r},\mathbf{r}^{\prime },i\omega ) &=&\sum_{\mathbf{k}j}%
\frac{\psi _{\mathbf{k}j\omega }^{R}(\mathbf{r})\psi _{\mathbf{k}j\omega
}^{L}(\mathbf{r}^{\prime })}{i\omega +\mu -E_{\mathbf{k}j\omega }}
\label{IMPeksGKS} \\
\mathcal{G}^{-1}(\mathbf{r},\mathbf{r}^{\prime },i\omega ) &=&\sum_{\mathbf{k%
}j}\psi _{\mathbf{k}j\omega }^{R}(\mathbf{r})(i\omega +\mu -E_{\mathbf{k}%
j\omega })\psi _{\mathbf{k}j\omega }^{L}(\mathbf{r}^{\prime })
\label{IMPeksGK1}
\end{eqnarray}%
\bigskip where the left $\psi _{\mathbf{k}j\omega }^{L}(\mathbf{r})$ and
right $\psi _{\mathbf{k}j\omega }^{R}(\mathbf{r})$ states satisfy to the
following Dyson equations:%
\begin{eqnarray}
\lbrack -\nabla ^{2}+V_{ext}(\mathbf{r})+V_{H}(\mathbf{r})]\psi _{\mathbf{k}%
j\omega }^{R}(\mathbf{r})+\int \mathcal{M}_{xc}(\mathbf{r},\mathbf{r}%
^{\prime },i\omega )\psi _{\mathbf{k}j\omega }^{R}(\mathbf{r}^{\prime })d%
\mathbf{r}^{\prime } &=&E_{\mathbf{k}j\omega }\psi _{\mathbf{k}j\omega }^{R}(%
\mathbf{r})  \label{IMPeksDER} \\
\lbrack -\nabla ^{2}+V_{ext}(\mathbf{r}^{\prime })+V_{H}(\mathbf{r}^{\prime
})]\psi _{\mathbf{k}j\omega }^{L}(\mathbf{r}^{\prime })+\int \psi _{\mathbf{k%
}j\omega }^{L}(\mathbf{r})\mathcal{M}_{xc}(\mathbf{r},\mathbf{r}^{\prime
},i\omega )d\mathbf{r} &=&E_{\mathbf{k}j\omega }\psi _{\mathbf{k}j\omega
}^{L}(\mathbf{r}^{\prime })  \label{IMPelsDEL}
\end{eqnarray}%
[we dropped the imaginary unit for simplicity in the notation $\psi _{%
\mathbf{k}j\omega }(\mathbf{r})$ which shall be thought as a shortened
version of $\psi _{\mathbf{k}j}(\mathbf{r},i\omega )$]. These equations
should be considered as the eigenvalue problems with complex non-hermitian
self--energy. As a result, the eigenvalues $E_{\mathbf{k}j\omega }$ [a
shortened form for $E_{\mathbf{k}j}(i\omega )$] being the same for both
equations are complex in general. The explicit dependency on the frequency $%
i\omega $ in both eigenvectors and eigenvalues comes from the self--energy.
Note that left and right eigenfunctions are orthonormal%
\begin{equation}
\int d\mathbf{r}\psi _{\mathbf{k}j\omega }^{L}(\mathbf{r})\psi _{\mathbf{k}%
j^{\prime }\omega }^{R}(\mathbf{r})\mathbf{=}\delta _{jj^{\prime }}
\label{IMPeksORT}
\end{equation}%
and can be used to evaluate the charge density of a given system using the
Matsubara sum and the integral over the k--space:%
\begin{equation}
\rho (\mathbf{r})=T\sum_{i\omega }\sum_{\mathbf{k}j}g_{\mathbf{k}j\omega
}\psi _{\mathbf{k}j\omega }^{L}(\mathbf{r})\psi _{\mathbf{k}j\omega }^{R}(%
\mathbf{r})e^{i\omega 0^{+}}  \label{IMPeksRHO}
\end{equation}%
where%
\begin{equation}
g_{\mathbf{k}j\omega }=\frac{1}{i\omega +\mu -E_{\mathbf{k}j\omega }}
\label{IMPeksOCC}
\end{equation}%
We \ have cast the notation of spectral density theory in a form similar to
DFT. The function $g_{\mathbf{k}j\omega }$ is the Green function in the
orthogonal left/right representation which plays a role of a "frequency
dependent occupation number".

It needs to be pointed out that the frequency dependent energy bands $E_{%
\mathbf{k}j\omega }$ represent an auxiliary set of complex eigenvalues.
These are not the true poles of the exact one--electron Green function $G(%
\mathbf{r},\mathbf{r}^{\prime },z)$ considered at complex z plane. However,
they are designed to reproduce the local spectral density of the system.
Note also that these bands $E_{\mathbf{k}jz}$ are not the true poles of the
auxiliary Green function $\mathcal{G}(\mathbf{r},\mathbf{r}^{\prime },z).$
The latter ones still need to be located by solving a non--linear equation
corresponding to the singularities in the expression (\ref{IMPeksGKS}) after
analytic continuation to real frequency. For a one--band case this equation
is simply: $z+\mu -E_{\mathbf{k}z}=0$ , whose solution delivers the
quasiparticle dispersion $Z_{\mathbf{k}}$. General knowledge of the poles
positions $Z_{\mathbf{k}j}$ will allow us to write an alternative expression
for $\mathcal{G}$ which is similar to (\ref{IMPeksGKS}), but with the
eigenvectors found at $Z_{\mathbf{k}j}$ thus carrying out no auxiliary
frequency dependence. These poles are the real one--electron excitational
spectra in case $\mathcal{G}$ is a good approximation to $G$. However, the
use of (\ref{IMPeksGKS}) is advantageous, since it avoids additional search
of poles and allows direct evaluation of the local spectral and charge
densities the system.

The energy--dependent representation allows us to obtain a very compact
expression for the total energy. As we have argued, the entropy terms are
more difficult to evaluate. However, they are generally small as long as we
stay at low temperatures. The pure kinetic part of the free energy expressed
via [see, Eq.(\ref{SDFactFRE})] 
\begin{equation}
Tr\ln \mathcal{G}-Tr\mathcal{M}_{eff}\mathcal{G}=T\sum_{i\omega }e^{i\omega
0^{+}}\int d\mathbf{r}d\mathbf{r}^{\prime }\ln \mathcal{G}(\mathbf{r},%
\mathbf{r}^{\prime },i\omega )-T\sum_{i\omega }\int d\mathbf{r}d\mathbf{r}%
^{\prime }\mathcal{M}_{eff}(\mathbf{r},\mathbf{r}^{\prime },i\omega )%
\mathcal{G}(\mathbf{r}^{\prime },\mathbf{r},i\omega )  \label{IMPeksKIN}
\end{equation}%
needs to be separated onto the energy and entropy terms. Both contributions
can be evaluated without a problem, but in light of neglecting the entropy
correction in the interaction part, we concentrate on evaluating \ the
kinetic energy only:%
\begin{equation}
T\sum_{i\omega }e^{i\omega 0^{+}}\int d\mathbf{r}\left[ (-\nabla _{\mathbf{r}%
}^{2})\mathcal{G}(\mathbf{r},\mathbf{r}^{\prime },i\omega )\right] _{\mathbf{%
r}=\mathbf{r}^{\prime }}=T\sum_{i\omega }e^{i\omega 0^{+}}\sum_{\mathbf{k}j}%
\frac{\langle \psi _{\mathbf{k}j\omega }^{L}|-\nabla ^{2}|\psi _{\mathbf{k}%
j\omega }^{R}\rangle }{i\omega +\mu -E_{\mathbf{k}j\omega }}
\label{IMPeksKI2}
\end{equation}%
The SDFT total energy formula is now arrived by utilizing the relationship $%
E_{\mathbf{k}j\omega }=\langle \psi _{\mathbf{k}j\omega }^{L}|-\nabla ^{2}+%
\mathcal{M}_{eff}|\psi _{\mathbf{k}j\omega }^{R}\rangle =\langle \psi _{%
\mathbf{k}j\omega }^{L}|-\nabla ^{2}+V_{ext}+V_{H}+\mathcal{M}_{xc}|\psi _{%
\mathbf{k}j\omega }^{R}\rangle $:

\begin{eqnarray}
&&E_{SDF}=T\sum_{i\omega }e^{i\omega 0^{+}}\sum_{\mathbf{k}j}g_{\mathbf{k}%
j\omega }E_{\mathbf{k}j\omega }-T\sum_{i\omega }\int d\mathbf{r}d\mathbf{r}%
^{\prime }\mathcal{M}_{eff}(\mathbf{r,r}^{\prime },i\omega )\mathcal{G}(%
\mathbf{r}^{\prime },\mathbf{r},i\omega )+  \notag \\
&&+\int d\mathbf{r}V_{ext}(\mathbf{r})\rho (\mathbf{r})+\frac{1}{2}\int d%
\mathbf{r}V_{H}(\mathbf{r})\rho (\mathbf{r})+\frac{1}{2}T\sum_{i\omega }\int
d\mathbf{r}d\mathbf{r}^{\prime }\mathcal{M}_{xc}(\mathbf{r,r}^{\prime
},i\omega )G_{loc}(\mathbf{r}^{\prime },\mathbf{r},i\omega )
\label{IMPeksSDF}
\end{eqnarray}%
If the self--energy is considered as input to the iteration while the Green
function is the output, near stationary point, it should have a convergency
faster than the convergency in the Green function.

It is instructive to consider the non--interactive limit when the
self--energy represents a local energy--independent potential, say, the
ground--state Kohn Sham potential of the density functional theory. This
provides an important test for our many-body calculation. It is trivial to
see that in the DFT limit, we obtain the Kohn--Sham eigenfunctions

\begin{eqnarray}
\psi _{\mathbf{k}j\omega }^{R}(\mathbf{r}) &\rightarrow &\psi _{\mathbf{k}j}(%
\mathbf{r})  \label{IMPeksPSL} \\
\psi _{\mathbf{k}j\omega }^{L}(\mathbf{r}) &\rightarrow &\psi _{\mathbf{k}%
j}^{\ast }(\mathbf{r})  \label{IMPeksPSR} \\
E_{\mathbf{k}j\omega } &\rightarrow &E_{\mathbf{k}j}  \label{IMPeksEKL}
\end{eqnarray}%
and the one--electron energy bands are no longer frequency dependent. The
sum over Matsubara frequencies in the expression for the charge density (\ref%
{IMPeksRHO}) can be performed analytically using the expression for the
Fermi--Diraq occupation numbers:%
\begin{equation}
f_{\mathbf{k}j}=\frac{1}{e^{(E_{\mathbf{k}j}-\mu )/T}+1}=T\sum_{i\omega }%
\frac{e^{i\omega 0^{+}}}{i\omega +\mu -E_{\mathbf{k}j}}  \label{IMPeksFKJ}
\end{equation}%
and the formula (\ref{IMPeksRHO}) collapses to the standard expression for
the density of non--interacting fermions. The total energy expression (\ref%
{IMPeksSDF}) is converted back to the DFT expression for the total energy
since the eigenvalue $E_{\mathbf{k}j\omega }$ becomes the DFT band structure 
$E_{\mathbf{k}j},$ and the summation over Matsubara frequencies $%
T\sum_{i\omega }e^{i\omega 0^{+}}g_{\mathbf{k}j\omega }$ gives according to (%
\ref{IMPeksFKJ}) the Fermi--Diraq occupation number $f_{\mathbf{k}j}$. The
standard DFT expression is recovered:

\begin{equation}
E_{DFT}=\sum_{\mathbf{k}j}f_{\mathbf{k}j}E_{\mathbf{k}j}-\int d\mathbf{r}%
V_{eff}(\mathbf{r})\rho (\mathbf{r})+\int d\mathbf{r}V_{ext}(\mathbf{r})\rho
(\mathbf{r})+\frac{1}{2}\int d\mathbf{r}V_{H}(\mathbf{r})\rho (\mathbf{r}%
)+E_{xc}[\rho ]  \label{IMPeksDFT}
\end{equation}%
where $E_{\mathbf{k}j}=\langle \psi _{\mathbf{k}j}|-\nabla ^{2}+V_{eff}|\psi
_{\mathbf{k}j}\rangle =\langle \psi _{\mathbf{k}j}|-\nabla
^{2}+V_{ext}+V_{H}+V_{xc}|\psi _{\mathbf{k}j}\rangle .$

\subsection{Use of Linear Muffin--Tin Orbitals}

The next problem is to solve the Dyson equation for the eigenvalues. The
sophisticated basis sets developed to solve the one--electron Schr\"{o}%
dinger equation can be directly used in this case. We utilize the LMTO
method described extensively in the past literature \cite%
{OKA-LMTO,TBLMTO,SavrasovLMTO} as it provides a minimal atom--centered local
orbital basis set ideally suited for the electronic structure calculation.\
Within the LMTO\ basis, the full Green function is represented as a sum%
\begin{equation}
\mathcal{G}(\mathbf{r},\mathbf{r}^{\prime },i\omega )=\sum_{\mathbf{k}%
}\sum_{\alpha \beta }\chi _{\alpha }^{\mathbf{k}}(\mathbf{r})\mathcal{G}%
_{\alpha \beta }(\mathbf{k},i\omega )\chi _{\beta }^{\mathbf{k\ast }}(%
\mathbf{r}^{\prime })  \label{IMPlmtGAB}
\end{equation}%
and, as we have argued in the previous section, the matrix $\mathcal{G}%
_{\alpha \beta }(\mathbf{k},i\omega )$ needs to be considered as a variable
in the spectral density functional. The stationarity yields the equation for
the Green function

\begin{equation}
\mathcal{G}_{\alpha \beta }(\mathbf{k},i\omega )=\left[ (i\omega +\mu )\hat{O%
}(\mathbf{k})-\hat{h}^{(0)}(\mathbf{k})-\mathcal{M}_{int}(\mathbf{k},i\omega
)\right] _{\alpha \beta }^{-1}  \label{IMPlmtINV}
\end{equation}%
where the overlap matrix $O_{\alpha \beta }(\mathbf{k})=\langle \chi
_{\alpha }^{\mathbf{k}}|\chi _{\beta }^{\mathbf{k}}\rangle ,$ the
non--interacting Hamiltonian matrix $h_{\alpha \beta }^{(0)}(\mathbf{k}%
)=\langle \chi _{\alpha }^{\mathbf{k}}|-\nabla ^{2}+V_{ext}(\mathbf{r})|\chi
_{\beta }^{\mathbf{k}}\rangle $ and the self--energy formally comes as a
matrix element%
\begin{equation}
\mathcal{M}_{int,\alpha \beta }(\mathbf{k},i\omega )=\int d\mathbf{r}d%
\mathbf{r}^{\prime }\chi _{\alpha }^{\mathbf{k\ast }}(\mathbf{r})\mathcal{M}%
_{int}(\mathbf{r},\mathbf{r}^{\prime },i\omega )\chi _{\beta }^{\mathbf{k}}(%
\mathbf{r}^{\prime })  \label{IMPlmtSIG}
\end{equation}%
over the LMTOs. Again, it is worth to point out that the self--energy here
depends on $\mathbf{k}$ via the orbitals even if the single--impurity case
is considered. In calculations performed on a cluster, the self--energy will
also pick its non--trivial $\mathbf{k}$--dependence coming from the nearest
sites.

While formally valid, the present approach is not very efficient since the
Green function $\mathcal{G}(\mathbf{r},\mathbf{r}^{\prime },i\omega )$ has
to be evaluated via (\ref{IMPlmtGAB}). This is the $\mathbf{k}$--integral
which has poles in a complex frequency plane, and integrating singular
functions need to be performed with care. In this respect, we adopt the
eigenvalue representation (\ref{IMPeksGKS}). We expand the energy--dependent
states in terms of the LMTO\ basis $\{\chi _{\alpha }^{\mathbf{k}}\}$ as
follows:

\begin{equation}
\psi _{\mathbf{k}j\omega }^{R(L)}(\mathbf{r})=\sum_{\alpha }A_{\alpha }^{%
\mathbf{k}j\omega ,R(L)}\chi _{\alpha }^{\mathbf{k}}(\mathbf{r})
\label{IMPlmtPSI}
\end{equation}%
The unknown coefficients $A_{\alpha }^{\mathbf{k}j\omega ,R(L)}$ are now the
quantities which have to be considered as variables in the spectral density
functional. The stationarity yields the equations%
\begin{eqnarray}
\sum_{\beta }\left[ h_{\alpha \beta }^{(0)}(\mathbf{k})+\mathcal{M}%
_{int,\alpha \beta }(\mathbf{k},i\omega )-E_{\mathbf{k}j\omega }O_{\alpha
\beta }(\mathbf{k})\right] A_{\beta }^{\mathbf{k}j\omega ,R} &=&0
\label{IMPlmtAAR} \\
\sum_{\alpha }A_{\alpha }^{\mathbf{k}j\omega ,L}\left[ h_{\alpha \beta
}^{(0)}(\mathbf{k})+\mathcal{M}_{int,\alpha \beta }(\mathbf{k},i\omega )-E_{%
\mathbf{k}j\omega }O_{\alpha \beta }(\mathbf{k})\right] &=&0
\label{IMPlmtAAL}
\end{eqnarray}%
These are the non--hermitian eigenvalue problems solved by standard
numerical methods. The orthogonality condition involving the overlap matrix
is%
\begin{equation}
\sum_{\alpha \beta }A_{\alpha }^{\mathbf{k}j\omega ,L}O_{\alpha \beta }(%
\mathbf{k})A_{\beta }^{\mathbf{k}j^{\prime }\omega ,R}=\delta _{jj^{\prime }}
\label{IMPlmtORT}
\end{equation}%
Note that the present algorithm just inverts the matrix (\ref{IMPlmtINV})
with help of the "right" and "left" eigenstates. The Green function (\ref%
{IMPlmtINV}) in the basis of its eigenvectors becomes

\begin{equation}
\mathcal{G}_{\alpha \beta }(\mathbf{k},i\omega )=\sum_{j}\frac{A_{\alpha }^{%
\mathbf{k}j\omega ,R}A_{\beta }^{\mathbf{k}j\omega ,L}}{i\omega +\mu -E_{%
\mathbf{k}j\omega }}  \label{IMPlmtEIG}
\end{equation}%
This formula can be safely used to compute the Green function as the
integral over the Brillouin zone, Eq. (\ref{IMPlmtGAB}), because the energy
denominator can be integrated analytically using the tetrahedron method \cite%
{Wigneron}.

Our next topic here is the evaluation of the bath Green function $\mathcal{G}%
_{0}(\mathbf{r},\mathbf{r}^{\prime },i\omega )$. It can be found from the
integral equation 
\begin{equation}
\mathcal{G}_{0}(\mathbf{r},\mathbf{r}^{\prime },i\omega )=G_{loc}(\mathbf{r},%
\mathbf{r}^{\prime },i\omega )-\int d\mathbf{r}^{\prime \prime }d\mathbf{r}%
^{\prime \prime \prime }G_{loc}(\mathbf{r},\mathbf{r}^{\prime \prime
},i\omega )\mathcal{M}_{int}(\mathbf{r}^{\prime \prime },\mathbf{r}^{\prime
\prime \prime },i\omega )\mathcal{G}_{0}(\mathbf{r}^{\prime \prime \prime },%
\mathbf{r},i\omega )  \label{IMPlmtDEQ}
\end{equation}%
where $\mathbf{r}$ and $\mathbf{r}^{\prime }$ run over $\Omega _{loc}$ In
order to solve this equation, it is useful to represent $\mathbf{r}=\mathbf{%
\rho }+\mathbf{R,}$ $\mathbf{r}^{\prime }=\mathbf{\rho }^{\prime }+\mathbf{R}%
^{\prime }\mathbf{,}$ and redenote $\mathcal{G}_{0}(\mathbf{r},\mathbf{r}%
^{\prime },i\omega )=\mathcal{G}_{0,RR^{\prime }}(\mathbf{\rho },\mathbf{%
\rho }^{\prime },i\omega ),G_{loc}(\mathbf{r},\mathbf{r}^{\prime },i\omega
)=G_{loc,RR^{\prime }}(\mathbf{\rho },\mathbf{\rho }^{\prime },i\omega ),%
\mathcal{M}_{int}(\mathbf{r},\mathbf{r}^{\prime },i\omega )=\mathcal{M}%
_{int,RR^{\prime }}(\mathbf{\rho },\mathbf{\rho }^{\prime },i\omega ).$
Considering one atom per unit cell let us see how this can be solved using
single--kappa LMTO--ASA method. The generalization to\ multiatomic systems
with multiple kappa basis sets as well as inclusion of full potential terms
in the calculation can be done along the same lines. The form of the LMTO\
basis function inside the sphere centered at $\mathbf{R}$ is%
\begin{equation}
\chi _{\alpha }^{\mathbf{k}}(\mathbf{r})=\chi _{\alpha }^{\mathbf{k}}(%
\mathbf{\rho })e^{i\mathbf{kR}}=e^{i\mathbf{kR}}[\Phi _{\alpha }^{H}(\mathbf{%
\rho })+\sum_{L}\Phi _{L}^{J}(\mathbf{\rho })S_{L\alpha }^{\mathbf{k}}]
\label{IMPlmtCHI}
\end{equation}%
where $S_{L\alpha }^{\mathbf{k}}$ are the structure constants of the LMTO$\ $%
method and where $\Phi _{L}^{H,J}(\mathbf{\rho })$ are linear combinations
of the solutions of the radial Schr\"{o}dinger equation taken at spherical
part of the potential and matched to the spherical Hankel (H) and Bessel (J)
functions at the muffin--tin sphere boundary as well as their energy
derivatives taken at some set of energies $E_{\nu l}$ at the center of
interest. The local Green function can be represented in this basis set as
follows%
\begin{equation}
G_{loc,RR^{\prime }}(\mathbf{\rho },\mathbf{\rho }^{\prime },i\omega
)=\sum_{LL^{\prime }}\sum_{\mu ,\nu =H,J}\Phi _{L}^{(\mu )}(\mathbf{\rho }%
)G_{loc,LRL^{\prime }R^{\prime }}^{(\mu \nu )}(i\omega )\Phi _{L^{\prime
}}^{(\nu )\ast }(\mathbf{\rho }^{\prime })  \label{IMPlmtGMN}
\end{equation}%
where the matrices $G_{loc,\alpha R\beta R^{\prime }}^{HH}(i\omega
),G_{loc,\alpha RL^{\prime }R^{\prime }}^{HJ}(i\omega ),G_{loc,LR\beta
R^{\prime }}^{JH}(i\omega ),G_{loc,LRL^{\prime }R^{\prime }}^{JJ}(i\omega )$
(indexes $R$ and $R^{\prime }$\ are restricted to a cluster) are given by
the following Brillouin zone integrals%
\begin{equation}
G_{loc,LRL^{\prime }R^{\prime }}^{(\mu \nu )}(i\omega )=\sum_{\mathbf{k}j}%
\frac{A_{L(\mu )}^{\mathbf{k}j\omega ,R}A_{L^{\prime }(\nu )}^{\mathbf{k}%
j\omega ,L}}{i\omega +\mu -E_{\mathbf{k}j\omega }}e^{i\mathbf{k}(\mathbf{R}-%
\mathbf{R}^{\prime })}  \label{IMPlmtGBZ}
\end{equation}%
\bigskip Here $A_{L(H)}^{\mathbf{k}j\omega ,R(L)}$ are the original
eigenvectors $A_{L}^{\mathbf{k}j\omega ,R(L)}$ and $A_{L(J)}^{\mathbf{k}%
j\omega ,L}=\sum_{\alpha }A_{\alpha }^{\mathbf{k}j\omega ,L}S_{L\alpha }^{%
\mathbf{k}},$ $A_{L(J)}^{\mathbf{k}j\omega ,R}=\sum_{\alpha }S_{L\alpha }^{%
\mathbf{k\ast }}A_{\alpha }^{\mathbf{k}j\omega ,R}$ are the convolutions of
the eigenvectors with the LMTO\ structure constants. We now utilize a
similar representation for the bath Green function 
\begin{equation}
\mathcal{G}_{0,RR^{\prime }}(\mathbf{\rho },\mathbf{\rho }^{\prime },i\omega
)=\sum_{LL^{\prime }}\sum_{\mu ,\nu =H,J}\Phi _{L}^{(\mu )}(\mathbf{\rho })%
\mathcal{G}_{0,LRL^{\prime }R^{\prime }}^{(\mu \nu )}(i\omega )\Phi
_{L^{\prime }}^{(\nu )\ast }(\mathbf{\rho }^{\prime })  \label{IMPlmtGRR}
\end{equation}%
where the matrices $\mathcal{G}_{0,\alpha R\beta R^{\prime }}^{HH}(i\omega ),%
\mathcal{G}_{0,\alpha RL^{\prime }R^{\prime }}^{HJ}(i\omega ),\mathcal{G}%
_{0,LR\beta R^{\prime }}^{JH}(i\omega ),\mathcal{G}_{0,LRL^{\prime
}R^{\prime }}^{JJ}(i\omega )$ can be found from the following \ Dyson
equation (where the matrices sizes have been enlarged by a factor of 2) 
\begin{equation}
\mathcal{G}_{0,LRL^{\prime }R^{\prime }}^{(\mu \nu )-1}(i\omega
)=G_{loc,LRL^{\prime }R^{\prime }}^{(\mu \nu )-1}(i\omega )+\mathcal{M}%
_{int,LRL^{\prime }R^{\prime }}^{(\mu \nu )}(i\omega )  \label{IMPlmtGLL}
\end{equation}%
with the self--energy matrices are defined as follows%
\begin{equation}
\mathcal{M}_{int,LRL^{\prime }R^{\prime }}^{(\mu \nu )}(i\omega )=\int d%
\mathbf{\rho }d\mathbf{\rho }^{\prime }\Phi _{L}^{(\mu )\ast }(\mathbf{\rho }%
)\mathcal{M}_{int,RR^{\prime }}(\mathbf{\rho },\mathbf{\rho }^{\prime
},i\omega )\Phi _{L^{\prime }}^{(\nu )}(\mathbf{\rho }^{\prime })
\label{IMPlmtSLL}
\end{equation}%
The solution of the impurity model with $\mathcal{G}_{0,LRL^{\prime
}R^{\prime }}^{(\mu \nu )}(i\omega )$ delivers new matrix elements (\ref%
{IMPlmtSLL}). The k--dependent self--energy (\ref{IMPlmtSIG}) to be used in
constructing the new Green function in (\ref{IMPlmtINV}) is found first by
restoring the $\mathbf{k}$--dependence from the cluster $\mathcal{M}%
_{int,LL^{\prime }}^{(\mu \nu )}(\mathbf{k},i\omega )=\sum_{R-R^{\prime }}%
\mathcal{M}_{int,LRL^{\prime }R^{\prime }}^{(\mu \nu )}(i\omega )e^{i\mathbf{%
k}(\mathbf{R}-\mathbf{R}^{\prime })}$ and second, restoring the $\mathbf{k}$%
--dependence of the LMTO\ basis as follows%
\begin{equation}
\mathcal{M}_{int,\alpha \beta }(\mathbf{k},i\omega )=\mathcal{M}_{int,\alpha
\beta }^{HH}(\mathbf{k},i\omega )+\sum_{L}\mathcal{M}_{int,\alpha L}^{HJ}(%
\mathbf{k},i\omega )S_{L\beta }^{\mathbf{k}}+\sum_{L^{\prime }}S_{L^{\prime
}\alpha }^{\mathbf{k\ast }}\mathcal{M}_{int,L^{\prime }\beta }^{JH}(\mathbf{k%
},i\omega )+\sum_{LL^{\prime }}S_{L\alpha }^{\mathbf{k\ast }}\mathcal{M}%
_{int,LL^{\prime }}^{JJ}(\mathbf{k},i\omega )S_{L^{\prime }\beta }^{\mathbf{k%
}}  \label{IMPlmtSAB}
\end{equation}

In practical calculations performed with the LDA+DMFT method for Pu, only
the subset of orbitals $\{\chi _{a}\}$ is treated as correlated (f electrons
of Pu) and a single--impurity case is considered. It is useful to separate
the Hartree and LDA exchange--correlations terms. Instead of dealing with
the non--interacting Hamiltonian in (\ref{IMPlmtINV}), we can rearrange the
contributions to arrive

\begin{equation}
\mathcal{G}_{\alpha \beta }(\mathbf{k},i\omega )=\left[ (i\omega +\mu )\hat{O%
}(\mathbf{k})-\hat{h}^{LDA}(\mathbf{k})-\Delta \mathcal{\hat{M}}(\mathbf{k}%
,i\omega )\right] _{\alpha \beta }^{-1}  \label{IMPlmtLDA}
\end{equation}%
where $h_{\alpha \beta }^{LDA}(\mathbf{k})=\langle \chi _{\alpha }^{\mathbf{k%
}}|-\nabla ^{2}+V_{LDA}(\mathbf{r})|\chi _{\beta }^{\mathbf{k}}\rangle $
with $V_{LDA}(\mathbf{r})=V_{ext}(\mathbf{r})+V_{H}(\mathbf{r})+V_{xc}^{LDA}(%
\mathbf{r}).$ The matrix $\Delta \mathcal{M}_{\alpha \beta }(\mathbf{k}%
,i\omega )=\mathcal{\tilde{M}}_{\alpha \beta }(\mathbf{k},i\omega
)-V_{\alpha \beta }^{DC}(\mathbf{k}),$ where $\mathcal{M}_{\alpha \beta }(%
\mathbf{k},i\omega )=\delta _{\alpha a}\delta _{\beta b}\mathcal{\tilde{M}}%
_{ab}(\mathbf{k},i\omega )$ and $V_{\alpha \beta }^{DC}(\mathbf{k})=\delta
_{\alpha a}\delta _{\beta b}V_{ab}^{DC}(\mathbf{k})$ represent the DMFT\
correction and double counting term described by (\ref{SDFlduSDC}). These
matrices are non--zero within the correlated subset. To accelerate the
calculation of the impurity model, we can parametrize the self--energy
matrix as $\mathcal{\tilde{M}}_{ab}^{(\mu \nu )}(i\omega )=\mathcal{\tilde{M}%
}_{ab}^{(p)}(i\omega )\langle \Phi _{a}^{(\mu )}|\Phi _{b}^{(\nu )}\rangle .$
With such parametrization, the local Green function which enters the Dyson
equation should be defined as follows $G_{loc,ab}(i\omega )=\sum_{\mathbf{k}%
j}\sum_{\mu \nu }A_{a(\mu )}^{\mathbf{k}j\omega ,R}$ $\langle \Phi
_{a}^{(\mu )}|\Phi _{b}^{(\nu )}\rangle A_{b(\nu )}^{\mathbf{k}j\omega ,L}/$ 
$(i\omega +\mu -E_{\mathbf{k}j\omega })$. This represents the generalization
of a partial--density--of--state formula of the LMTO method. The bath Green
function can be found from the equation: $\mathcal{G}_{0,ab}(i\omega
)^{-1}=G_{loc,ab}(i\omega )^{-1}+\mathcal{\tilde{M}}_{ab}^{(p)}(i\omega )$
and can be passed to the impurity solver. The latter delivers a new
self--energy $\mathcal{\tilde{M}}_{ab}^{(p)}(i\omega )$ which is then
multiplied by $\langle \Phi _{a}^{(\mu )}|\Phi _{b}^{(\nu )}\rangle $ and
used to reconstruct new $\mathbf{k}$--dependent self--energy after (\ref%
{IMPlmtSAB}). Such procedure preserves all $\mathbf{k}$--dependent
information coming from the orbitals.

\section{Applications to Plutonium}

This section describes the application of the theory to Plutonium. Pu is
known to be an anomalous metal \cite{PuBook}. It has six crystallographic
structures. Starting from the low temperature $\alpha $ phase (0 to 100 C)
with 16 atoms per unit cell it shows a series of phase transitions and ends
up in relatively simple fcc $\delta $ (300--450 C) and bcc $\varepsilon $
phases (500--650 C) just before it melts. The temperature dependence of
atomic volume in Pu is anomalous \cite{PuVol}. It shows an enormous volume
expansion between $\alpha $ and $\delta $ phases which is about 25\%. Within
the $\delta $ phase, the metal shows negative thermal expansion. Transition
between $\delta $ and higher--temperature $\varepsilon $ phase occurs with a
5\% volume collapse. Also, Pu shows anomalous resistivity behavior \cite%
{PuRHO} characteristic for the heavy fermion systems, but neither of its
phases is magnetic. The susceptibility is small and relatively temperature
independent. The photoemission \cite{PuExp1} shows a strong narrow
Kondo--like peak at the Fermi level consistent with large values of the
linear specific heat coefficient.

Density functional based LDA\ and GGA\ calculations describe the properties
of Pu incorrectly. They predict magnetic ordering \cite{PuMag}. They
underestimate \cite{PuLDA} equilibrium volume of the $\delta $ and $%
\varepsilon $ phase by as much as 30\% , the largest discrepancy known in
LDA, which usually predicts the volume of solids within a few \% accuracy
even for such correlated systems as high temperature superconductors.
Despite this, the volume of the $\alpha $ phase is predicted correctly by
LDA \cite{PuLDA}. Since the transport and thermodynamic properties of $%
\alpha $ and $\delta $ Pu are very similar, the nature of the $\alpha $
phase and the reason why LDA predicts accurately its structure and volume is
by itself is another puzzle.

To address these questions several approaches have been developed. The LDA+U
method was applied to $\delta $ Pu \cite{PuPRL,PuLDA+U}. It is able to
produce the correct volume of the $\delta $ phase for values of the
parameter $U\symbol{126}$4 eV consistent with atomic spectral data and
constrained density functional calculations. Similar calculation has been
performed by a so--called orbitally ordered density functional method \cite%
{PuLDA+J}. However, both methods predict Pu to be magnetic, which is not
observed experimentally. The LDA+U method is unable to predict the correct
excitation spectrum. Also, to recover the $\alpha $ phase within LDA+U$\ $%
the parameter U has to be set to zero which is inconsistent with its
transport properties and with microscopic calculations of this parameter.
Another approach proposed \cite{PuCore} in the past is the constrained LDA
approach in which some of the 5f electrons, are treated as core, while the
remaining are allowed to participate in band formation. Results of the
self--interaction--corrected LDA calculations have been reported \cite{PuSIC}%
, as well as qualitative discussion of the bonding nature across the
actinides series has been given \cite{Harrison}.

Thus, the problem of Pu is challenging because its f-electrons are close to
the Mott transition \cite{Johansson}. It provides us a crucial test for our
quantitative theory of strong correlations. A short version of this work has
appeared already \cite{PuNature}. Our implementation is based on the
self--consistent LDA+DMFT method and uses the LMTO method in its
tight-binding representation \cite{TBLMTO}. Spin--orbit coupling effects are
important for actinide compounds and have been included in the calculation
for Pu. The \textquotedblleft full potential\textquotedblright\ terms have
been neglected in the calculation through the use of the atomic sphere
approximation with a one-kappa LMTO basis set. The necessary $\mathbf{k}$%
-space integrals for evaluating Green functions and charge densities have
been carried out using the tetrahedron method using (8,8,8) grid in the
Brillouin zone.

We study in detail the total energy as a function of the parameter $U$ and
give our predictions for the volumes in $\alpha ,$ $\delta ,$ and $%
\varepsilon $ phases. A comparative analysis of the one--electron spectra in
both $\alpha $ and $\delta $ phases is also given and the comparison with
the photoemission experiment \cite{PuExp1} is made. Since the dynamical mean
field theory requires the solution of the Anderson impurity model for the
multi-orbital f--shell of Pu, we have developed a method which, inspired by
the success of the iterative perturbation theory \cite{ReviewDMFT},
interpolates the self--energy between small and large frequencies. At low
frequencies, the exact value of the self-energy and its slope is extracted
from the Friedel sum rule and from a slave-boson treatment\cite%
{Gutz,Ruck,Fleszar}. This approach is accurate as it has been shown recently
to give the exact critical value of $U$ in the large degeneracy limit \cite%
{Parcollet} \ At high frequencies the self-energy behavior can be computed
based on high--frequency moments expansions\cite{Hubbard1,Moments}. The
result of interpolation can be encoded into a simple form for the
self--energy like the following continuous fraction expansion%
\begin{equation}
\Sigma (i\omega )=\Sigma (i\infty )+\frac{A}{i\omega -\frac{B}{i\omega -C}}
\label{APLimpSIG}
\end{equation}%
where the unknown coefficients are determined to satisfy known conditions in
the low-- and high--frequency limits. This kind of self--energy fits Quantum
Monte Carlo (QMC) data in large region of parameters, such as $U$ and
doping, and where this comparison is at all possible (small degeneracy and
high temperature). Thus, our approach interpolates between four major
limits: small and large i$\omega $'s valid for any $U$ as well as small and
large $U^{\prime }s$ (band vs. atom) valid for any $i\omega $.\ The
analytical continuation to the real frequency axis is not a problem with the
present method and avoids the use of the Pade \cite{Pade} and maximum
entropy \cite{ReviewQMC} based techniques.

Complete details of the method can be found in Ref. \onlinecite{Udo}. Here
we only mention technicalities connected to the f--electrons of Pu where we
deal with the impurity Green functions which are the matrices 14x14.
However, for the relativistic f-level in cubic symmetry, the matrices can be
reduced to 5x5 with 4 non-zero off--diagonal elements. The solution of such
impurity problem is still a formidable numerical problem. We therefore make
some simplifications. First, the off--diagonal elements are in general small
and will be neglected. We are left with the 5f$^{5/2}$ state split into 2
levels which are 2-fold ($\Gamma _{7}$), and 4-fold ($\Gamma _{8}$)
degenerate, and with the 5f$^{7/2}$ state split into 3 levels which are
2-fold ($\Gamma _{6}$), 2-fold ($\Gamma _{7}$) and 4-fold ($\Gamma _{8}$)
degenerate. Second, since in Pu the intermultiplet spin-orbit splitting is
much larger than the intramultiplet crystal field splitting (\TEXTsymbol{>}
5:1), we reduce the problem of solving AIM for the levels separately by
treating the 5f$^{5/2}$ $\Gamma _{7}$ and $\Gamma _{8}$ levels as one 6-fold
degenerate level, and the 5f$^{7/2}$ $\Gamma _{6}$, $\Gamma _{7}$ and $%
\Gamma _{8}$ levels as another 8-fold degenerate level.

\subsection{Calculation of Volume}

We perform our calculations for $\delta $ and $\epsilon $ phases of Pu
having simple fcc and bcc structures respectively. Since our method does not
yet allow us to treat complicated lattices, only a simplified study of the $%
\alpha $ phase which formally has 16 atoms per unit--cell will be reported.\
The total energy as a function of volume is evaluated self--consistently
using formula (\ref{IMPeksSDF}). It contains all the electrons including the
core electrons. The local density approximation includes generalized
gradient corrections after Ref. \onlinecite{Perdew}. Since the LDA+DMFT
approximation (\ref{SDFldaLDA}) is used, we subtract from the LDA the
average interaction energy of the f electrons in the form (\ref{SDFlduFDC})
of the double counting term and then add improved estimates of these
quantities using the self--consistent solution of the impurity model.

To illustrate the importance of correlations, we discuss the results for
various strengths of the on--site Coulomb interaction $U$. Fig. \ref%
{FigTotalEnergy} reports our theoretical predictions. First, the total
energy as a function of volume of the fcc lattice is computed. The
temperature is fixed at $600K$, i.e. in the vicinity of the region where the 
$\delta $ phase is stable. $U$=0 GGA curve indicates a minimum at $V/V_{0}$%
=0.7. This volume is in fact close to the volume of the $\alpha $ phase.
Certainly, we expect that correlations should be less important for the
compressed lattice in general, but there is no sign whatever of the $\delta $
phase in the $U$=0 calculation. The total energy curve is dramatically
different for $U$\TEXTsymbol{>}0. The details depend sensitively on the
actual value of $U$. The behavior at $U$=4 eV shows the possibility of a
double minimum; it is actually realized for a slightly smaller value of $U$.
We find that for $U=$ 3.8 eV, the minimum occurs near $V/V_{\delta }$=0.80
which corresponds to the volume of the $\alpha $ phase if we allow for
monoclinic distortions and a volume--dependent $U$. When $U$ increases by
0.2 eV the minimum occurs at $V/V_{\delta }$=1.05 which corresponds to the
volume of the $\delta $ phase, in close agreement with experiment. Since the
energies are so similar, we may expect that as temperature decreases, the
lattice undergoes a phase transition from the $\delta $ phase to the $\alpha 
$ phase with the remarkable decrease of the volume by 25\%.

\begin{figure}[tbh]
\includegraphics*[height=3.0in]{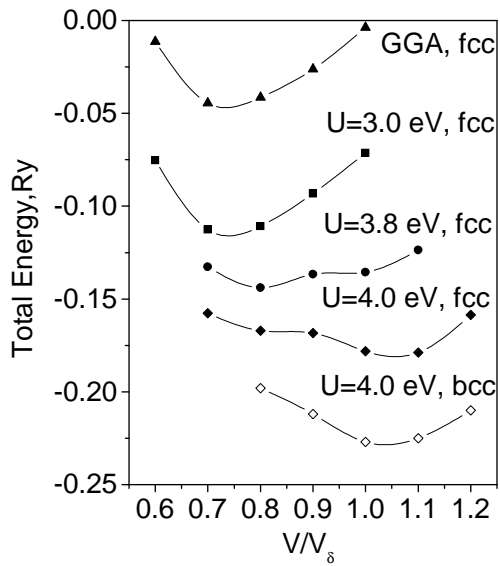}
\caption{Total energy as a function of volume in Pu for different values of $%
U$ calculated using the LDA+DMFT approach. Data for the fcc lattice are
computed at T=600K, while data for the bcc lattice are given for T=900K.}
\label{FigTotalEnergy}
\end{figure}

We repeated our calculations for the bcc structure using the temperature $%
T=900K$ where the $\epsilon $ phase is stable. Fig. \ref{FigTotalEnergy}
shows these results for $U=$ 4 eV with a location of the minimum at around $%
V/V_{\delta }$=1.03. While the theory has a residual inaccuracy in
determining the $\delta $ and $\epsilon $ phase volumes by a few percent, a
hint of volume decrease with the $\delta \rightarrow \epsilon $ transition
is clearly reproduced. Thus, our first-principles calculations reproduce the
main features of the experimental phase diagram of Pu.

Note that the values of $U\sim $4 eV which are needed in our simulation to
describe the $\alpha \rightarrow \delta $ transition, are in good agreement
with the values of on--site Coulomb repulsion between f--electrons estimated
by atomic spectral data \cite{PuU1}, constrained density functional studies 
\cite{PuU2}, and our previous LDA+U studies\cite{PuPRL}.

The double-well behavior in the total energy curve is unprecedented in LDA
or GGA based calculations but it is a natural consequence of the proximity
to a Mott transition. Indeed, recent studies of model Hamiltonian systems 
\cite{ReviewDMFT,Sahana} in the vicinity of the Mott transition show that
two DMFT solutions which differ in their spectral distributions can coexist.
It is very natural that allowing the density to relax in these conditions
can give rise to the double minima as seen in Fig. \ref{FigTotalEnergy}.

\subsection{Calculation of Spectra}

\begin{figure}[tbh]
\includegraphics*[height=3.0in]{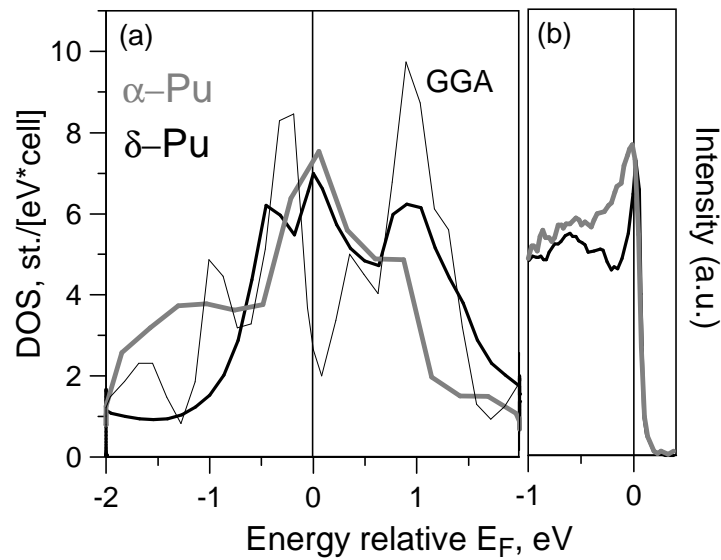}
\caption{a) Comparison between calculated densities of states using the
LDA+DMFT approach for fcc Pu: the data for $V/V_{\protect\delta %
}=1.05,\,U=4.0$ eV (black line) , the data for $V/V_{\protect\delta %
}=0.80,\,\,U=3.8$ eV (gray line) which correspond to the volumes of the $%
\protect\delta $ and $\protect\alpha $ phases respectively. The result of
the GGA\ calculation (dotted line) at $V/V_{\protect\delta }=1\,\,(U=0)$ is
also given. b) Measured photemission spectrum of $\protect\delta $ (black
line) and $\protect\alpha $ (grey line) Pu at the scale from -1.0 to 0.4 eV
(after Ref. \onlinecite{PuExp1}).}
\label{FigSpectra}
\end{figure}

We now report our calculated spectral density of states for the fcc
structure using the volume $V/V_{\delta }$=0.8 and $V/V_{\delta }$=1.05
corresponding to our theoretical studies of $\alpha $ and $\delta $ phases.
To compare the results of the dynamical mean--field calculations with the
LDA method as well as with the experiment, we discuss the results presented
in Fig. \ref{FigSpectra}. Fig. \ref{FigSpectra}(a) shows the density of
states calculated using LDA+DMFT method in the vicinity of the Fermi level.
Solid black line corresponds to the $\delta $ phase and solid grey line
corresponds to the $\alpha $ phase. We predict the appearance of a strong
quasiparticle peak near the Fermi level which exists in the both phases.
Also, the lower and upper Hubbard bands can be clearly distinguished in this
plot. The width of the quasiparticle peak in the $\alpha $ phase is found to
be larger by 30 per cent compared to the width in the $\delta $ phase. This
indicates that the low-temperature phase is more metallic, i.e. it has
larger spectral weight in the quasiparticle peak and smaller weight in the
Hubbard bands. Recent advances have allowed the experimental determination
of these spectra, and our calculations are consistent with these
measurements \cite{PuExp1}. Fig. \ref{FigSpectra}(b) shows the measured
photoemission spectrum for $\delta $ (black line) and $\alpha $ (gray line)
Pu. We can clearly see a strong quasiparticle peak. Also a smaller peak
located at 0.8 eV for the $\delta $--phase can be found. We interpret it as
the lower Hubbard band.

The result of the local density approximation is shown on Fig. \ref%
{FigSpectra}(a) by dashed line. The LDA produces two peaks near the Fermi
level corresponding to 5f$^{5/2}$ and 5f$^{7/2}$ states separated by the
spin-orbit coupling. The Fermi level falls into the dip between these states
and cannot reproduce the features seen in photoemission. We should also
mention that LDA+U\ fails completely \cite{PuPRL,PuLDA+U} to reproduce the
intensity of the $f$--states near the Fermi level as it pushes the $f$--band
2--3 eV below the Fermi energy. This is the picture expected from the static
Hartree--Fock theory such as the LDA+U. Only full inclusion of the dynamic
effects within the DMFT allows to account for both the quasiparticle
resonance and the Hubbard satellites which explains all features of the
photoemission spectrum in $\delta $ Pu.

The calculated by LDA+DMFT densities of states at E$_{F}$ equal to 7
st./[eV*cell] are consistent with the measured values of the linear specific
heat coefficient. We still find a residual discrepancy by about factor of 2
due to either inaccuracies of the present calculation or due to the
electron--phonon interactions. However, these values represent an
improvement as compared to the LDA calculations which appear to be 5 times
smaller. Similar inaccuracy has been seen in the LDA+U calculation \cite%
{PuPRL}.

A simple physical explanation drawn from these studies suggests that in the $%
\delta $ phase the f electrons are slightly on the localized side of the
interaction--driven localization-delocalization transition with a sharp and
narrow Kondo--like peak and well-defined upper and lower Hubbard bands. It
therefore has the largest volume as has been found by previous LDA+U
calculations\cite{PuPRL,PuLDA+U} which take into account Hubbard bands only.
The low-temperature $\alpha $ phase is more metallic, i.e. it has larger
spectral weight in the quasiparticle peak and smaller weight in the Hubbard
bands. It will therefore have a much smaller volume that is eventually
reproduced by LDA/GGA calculations which neglect both Coulomb
renormalizations of quasiparticles and atomic multiplet structure. The
delicate balance of the energies of the two minima may be the key to
understanding the anomalous properties of Pu such as the great sensitivity
to small amounts of impurities (which intuitively would raise the energy of
the less symmetric monoclinic structure, thus stabilizing the $\delta $
phase to lower temperature) and the negative thermal expansion. Notice
however, that the $\alpha $ phase is not a weakly correlated phase: it is
just slightly \ displaced towards the delocalized side of the
localization--delocalization transition, relative to the $\delta $ phase.
This is a radical new viewpoint in the theoretical literature on Pu, which
has traditionally regarded the $\alpha $ phase as well understood within
LDA. However, the correlation viewpoint is consistent with a series of
anomalous transport properties in the $\alpha $ phase reminiscent of heavy
electron systems. For example, the resistivity of $\alpha $--Pu around room
temperature is anomalously large, temperature independent and above the Mott
limit \cite{PuRHO} (the maximum resistivity allowed to the conventional
metal). Strong correlation anomalies are also evident in the thermoelectric
power \cite{PuTHERM}.

\section{Conclusion}

In conclusion, this work describes a first principles method for calculating
the electronic structure of materials where many--body correlation effects
between the electrons are not small and cannot be neglected. It allows
simultaneous evaluation of the total free energy and the local electronic
spectral density. The approach is based on the effective action functional
formulation of the free energy and is viewed as spectral density functional
theory. An approximate form of the functional exploits a local dynamical
mean--field theory of strongly correlated systems accurate in the situations
when the self--energy is short ranged in a certain portion of space. The
localization is defined with reference to some basis in Hilbert space. It
does not necessarily imply localization in real space and is treated using a
general basis set following the ideology of the cellular dynamical mean
field theory. Further approximations of the theory, such as LDA+DMFT and
local GW are discussed. Implementation of the method is described in terms
of the energy--dependent one--particle states expanded via the linear
muffin--tin orbitals. Application of the method in its LDA+DMFT form is
given to study the anomalous volume expansion in metallic Plutonium. We
obtain equilibrium volume of the $\delta $ phase in good agreement with
experiment with no magnetic order imposed in the calculation. The calculated
one--electron densities of states are consistent with the results of the
photoemission. Our most recent studies \cite{PuScience} of the lattice
dynamical properties of Pu address the problem of the $\delta \rightarrow
\varepsilon $ transitions and show good agreement with experiment\cite%
{PuWong}.

Alternative developments of the LDA+DMFT approach by several groups around
the world discuss other applications of the dynamical mean field theory in
electronic structure calculations. The results obtained are promising.
Volume collapse transitions, materials near the Mott transition, systems
with itinerant and local moments, as well as many other exciting problems
are beginning to be explored using these methods.

The authors would like to thank for many enlightening discussions the
participants of the research school on Realistic Theories of Correlated
Electron Materials, Kavli Institute for Theoretical Physics, fall 2002,
where the part of this work was carried out. Many helpful discussions with
the participants of weekly condensed matter seminar at Rutgers University
are also acknowledged. Conversations with A. Georges, A Lichtenistein, N. E.
Zein related to various aspects of dynamical mean--field and GW theories are
much appreciated. The authors are indebted to E. Abrahams, A. Arko, J.
Joyce, H. Ledbetter, A. Migliori, and J. Thompson for discussing the issues
related to the work on Pu. The work was supported by the NSF DMR Grants No.
0096462, 02382188, 0312478, 0342290, \ US DOE division of Basic Energy
Sciences Grant No DE-FG02-99ER45761, and by Los Alamos National Laboratory
subcontract No 44047-001-0237. Kavli Institute for Theoretical Physics is
supported by NSF grant No. PHY99-07949.



\end{document}